\documentclass[final, 5p, times, twocolumn]{elsarticle}

\usepackage{amsmath}  
\usepackage{amssymb}
\usepackage{algorithm}
\usepackage{makecell}
\usepackage{graphicx}
\usepackage{balance}
\usepackage{caption}
\usepackage{subcaption}
\usepackage{subfloat}
\usepackage{comment}
\usepackage{algorithm}
\usepackage{algpseudocode}
\usepackage[utf8]{inputenc}
\usepackage{makecell}
\usepackage{kotex}
\usepackage{tabularx}
\usepackage{booktabs}
\usepackage{multirow}
\usepackage[hidelinks]{hyperref}
\usepackage{xurl}
\usepackage{natbib}
\usepackage{etoolbox}
\usepackage{microtype}
\usepackage{enumitem}

\begin{document}

\begin{frontmatter}

\title{SpotVista: Availability-Aware Recommendation System for Reliable and Cost-Efficient Multi-Node Spot Instances}

\author[hyu-ds]{Taeyoon Kim}
\author[hyu-ds]{Kyumin Kim}
\author[kt]{Kyunghwan Kim}
\author[hyu-ds]{Hayoung Kim}
\author[hyu-ds]{Seungwoo Jeong}
\author[hyu-ai]{Moohyun Song}
\author[hyu-ds]{Kyungyong Lee\corref{cor1}}
\ead{kyungyong@hanyang.ac.kr}
\cortext[cor1]{Corresponding author}
\affiliation[hyu-ds]{organization={Hanyang University},
            addressline={Department of Data Science}, 
            city={Seoul},
            postcode={04763}, 
            country={Republic of Korea}}
\affiliation[kt]{organization={KT}, 
            city={Seoul},
            postcode={06763}, 
            country={Republic of Korea}}
\affiliation[hyu-ai]{organization={Hanyang University},
            addressline={Department of Artificial Intelligence}, 
            city={Seoul},
            postcode={04763}, 
            country={Republic of Korea}}

\begin{abstract}
Cloud vendors offer discounted spot instances to maximize surplus resource utilization, but these instances are subject to the risk of sudden interruption. Traditional pricing datasets have been employed to predict this risk, yet recent policy changes by cloud vendors have diminished their effectiveness. To promote spot instance usage, public cloud vendors provide instant availability datasets to help users mitigate interruption risks. While existing research utilizing this data has proposed methods to reduce interruptions, these studies have primarily focused on single-node instances, overlooking the stability of multi-node environments widely adopted for modern cloud workloads.

This paper proposes SpotVista, a system that recommends a resource pool of reliable and cost-efficient multi-node spot instances by leveraging various publicly available datasets. To achieve this, SpotVista collects a large-scale multi-node availability dataset while overcoming significant query limitations. Through a thorough analysis of multi-node spot instance availability behavior, SpotVista establishes a methodology for recommending cost-efficient and reliable multi-node configurations.

To evaluate how effectively the proposed methodology reflects multi-node availability and cost efficiency, extensive real-world interruption experiments were conducted. The results demonstrate that SpotVista outperforms the state-of-the-art work, SpotVerse, achieving 81.28\% greater availability and 2.84\% more cost savings in a multi-region setup. When compared to a publicly available service, AWS SpotFleet, SpotVista provides 21.6\% higher stability and 26.3\% greater cost savings.
\end{abstract}


\end{frontmatter}

\section{Introduction}
Cloud computing has fundamentally altered the way of consuming computing resources by offering elastic capacity without the overhead of managing physical infrastructure. To provide users with such flexibility, cloud service vendors maintain resources beyond peak demand, which results in a surplus. To maximize the utilization of these otherwise idle resources, cloud vendors offer them at significant discounts of up to 90\%, referring to them as spot instances. This service model has been adopted by most major cloud vendors, including AWS, Azure, GCP, Alibaba, and IBM.

Initially, the price of spot instances was dynamically adjusted based on supply and demand, and an instance may be reclaimed, an event known as spot instance interruption, when the market price exceeds a user's bid price. Instance interruptions pose a significant threat to application reliability, requiring users to prepare a mechanism to deal with such events. To assist users in this endeavor, cloud vendors have historically provided public datasets, most notably spot pricing data. This dataset was extensively used in prior research to analyze price behavior with the interruption events~\cite{deconstructing-spot-instance, alibaba-spot-instance, spot-analysis-javadi,spot-instance-analysis, spot-price-by-location}, enhance application reliability~\cite{deepspotcloud, see-spot-run, flint, spotweb, spot-for-online-service, spot-batch, spoton, autobot-bot-using-spot}, and predict optimal bid prices to minimize interruption risk~\cite{how-to-bid-cloud, how-not-to-bid-cloud, spot-price-prediction, spot-price-rf, spot-bidding-infocom, spot-price-prediction-dnn}. However, recent changes in cloud vendor policies have weakened the correlation between spot prices and instance terminations, reducing the effectiveness of these price-based predictive models~\cite{spot-price-change-2017, spot-price-policy-change-2017-irwin}.

In response to this shift, cloud vendors have introduced new availability-focused datasets, such as AWS and Azure Spot Placement Score (SPS), which provide real-time indicators of spot instance availability~\cite{spot-placement-score-start, azure-spot-placement-score}. While services like SpotLake have emerged to archive and provide access to these datasets~\cite{spotlake-iiswc, multi-spotlake-www}, they are limited in scope. Specifically, they offer availability information only for single-node instances and do not adequately address the requirements of modern distributed applications, such as large-scale machine learning model training and big data processing, which often depend on multi-node environments. Cheon et al.~\cite{multinode-spot-dataset} presented the discrepancy of a single node SPS dataset when inferring availability of multi-node spot instances pool, and they emphasized the inadequacy of single-node metrics for provisioning reliable, large-scale resource pools. Furthermore, the dataset archive service provides only raw dataset, and users must choose cost-efficient and reliable spot instances manually that can be a very challenging task. Although cloud vendors offer native recommendation services like AWS SpotFleet, their internal operational mechanisms are not disclosed, and the effectiveness of their recommendations has not been independently validated.

To address these limitations by exploring the availability characteristics of multi-node spot instance environments, we propose a heuristic to efficiently collect a comprehensive multi-node availability dataset followed by performing thorough analysis for the dataset. Based on this dataset, this work introduces a novel availability score to quantify the stability of multi-node spot instance configurations. This scoring mechanism serves as the foundation for a recommendation algorithm designed to identify spot instance pools that achieve an optimal balance between cost-efficiency and high availability.

The effectiveness of the proposed system is validated through extensive experiments on real-world cloud infrastructure, involving 59,560 spot requests across 127 unique instance types. The proposed algorithm demonstrates 81.28\% higher availability and 2.84\% greater cost savings than the state-of-the-art system, SpotVerse~\cite{spotverse}, and provides 21.6\% higher availability with 26.3\% more cost savings compared to the commercial AWS SpotFleet service. The collected multi-node availability dataset and the recommendation engine are made publicly available as a web service to benefit the research community and cloud practitioners\footnote{https://spotvista.ddps.cloud}.

The main contributions of this paper are as follows:
\begin{itemize}[noitemsep, topsep=0em, labelsep=0.5em]
    \item The development of a novel sampling heuristic for the efficient collection of a comprehensive multi-node spot instance availability dataset, addressing significant query limitations.
    \item A detailed analysis of the collected dataset, revealing key temporal and spatial characteristics of multi-node spot instance availability.
    \item The design and validation of a spot instance recommendation engine that quantifies both availability and cost-efficiency, whose effectiveness is demonstrated through extensive real-world experiments.
    \item The deployment of a publicly available web service that provides both the collected multi-node availability data and the recommendation engine.
\end{itemize}

\section{Spot Instance and Availability}
\begin{figure}[t]
    \centering
    \includegraphics[width=\columnwidth]{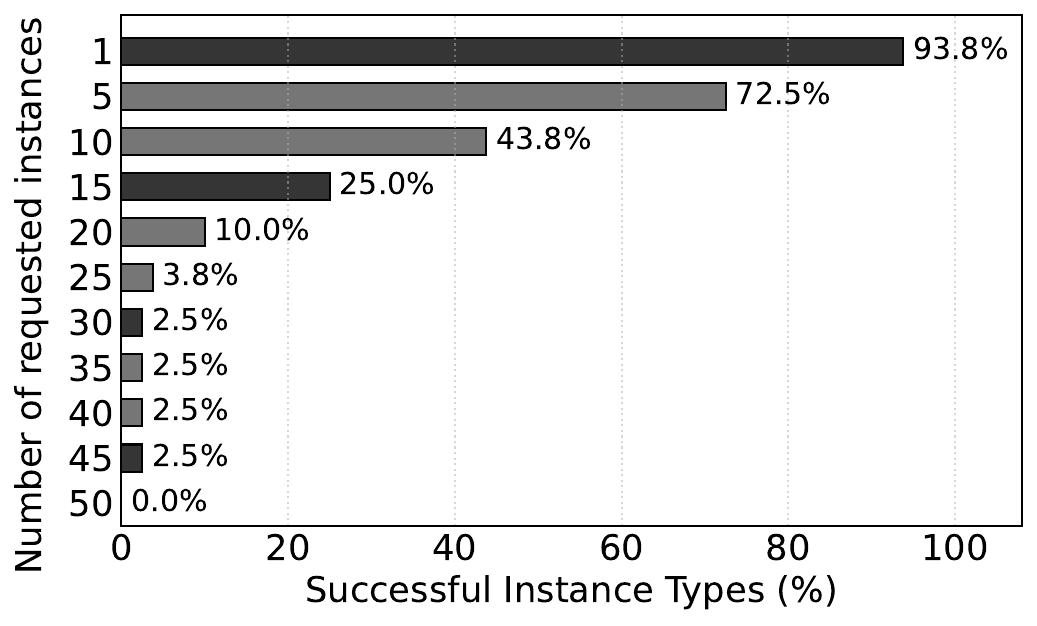}
    \caption{Success rate of spot instance allocation for types with a single-node SPS of 3. Each type was requested 50 times, and the bars indicate the fraction of types that achieved at least the corresponding number of successful allocations.}
    \label{fig:sps-score-distribution-one-fifty-nodes}
\end{figure}

For spot instance users, balancing cost savings with operational stability is a primary consideration, given the dynamically changing prices and the inherent risk of interruption events. To assist users, public cloud vendors offer various datasets that are related to spot instance prices and stability. In the past, particularly within auction-based models (e.g., AWS~\cite{optimal-spot-bidding-cloud}, Azure~\cite{azure-bidding}, Alibaba~\cite{alibaba-bidding}), these datasets were widely employed to estimate interruption risks~\cite{how-to-bid-cloud, how-not-to-bid-cloud}. For instance, AWS provided regular updates and historical archives of spot prices, which facilitated a significant body of research on their stable usage~\cite{deconstructing-spot-instance, spot-analysis-javadi, spot-price-prediction}. However, recent policy changes have diminished the correlation between spot prices and interruptions. The reduced frequency of price updates has consequently limited the effectiveness of the previous research~\cite{spot-price-policy-change-2017-irwin, spot-price-change-2017}.

In response, cloud vendors have shifted to providing datasets directly related to spot instance availability, and AWS and Azure have started to offer metrics such as interruption ratios over the past month and real-time availability data, exemplified by SPS~\cite{spot-placement-score-start, azure-spot-placement-score}. Although the internal calculation details are not disclosed, these metrics are described as indicators of immediate spot instance availability. For example, AWS's SPS assigns a score of $1$ (Low), $2$ (Medium), or $3$ (High) to each instance type, where a higher score signifies greater availability. While users can access these datasets through management consoles or APIs, they are often constrained by query limitations. To overcome these constraints and facilitate easier data access, a web service named SpotLake has been developed~\cite{spotlake-iiswc, multi-spotlake-www}.

However, a critical limitation of existing availability datasets and services is their focus on single-node instances that are misaligned with the demands of modern cloud applications. Workloads such as large-scale deep learning model training~\cite{deep-speed-inference} and distributed big data processing~\cite{lithops-tcc} inherently require multi-node environments and incur significant computational costs. Consequently, leveraging spot instances to mitigate these costs has become a compelling strategy for distributed systems, leading to a substantial body of research. This research spans various domains, including DNN training~\cite{bamboo-nsdi, parcae-nsdi}, big data processing~\cite{spot-mapreduce-hotcloud}, and general-purpose web services~\cite{spotweb, spot-batch, autobot-bot-using-spot}.

The majority of this existing work concentrates on reactively handling interruptions after an instance has been reclaimed. Some studies have explored proactively selecting low-risk instances to prevent interruptions. For example, Kim et al. proposed a method to enhance stability by predicting SPS values~\cite{interrupt-visible-www}, and SpotVerse~\cite{spotverse} references SPS and interruption frequency score to infer spot instance availability. Yet, these approaches, like others, are constrained by their reliance on single-node spot instance data, limiting their applicability to distributed environments. For large-scale multi-node deployments, an availability metric that reflects the stability of an entire batch of many instances is essential.

Cheon et al.~\cite{multinode-spot-dataset} examined the inadequacy of single-node SPS for predicting availability in multi-instance requests. Their experiments showed that while a single spot instance request succeeded in all trials, the success rate decreased substantially as the requested count increased. For a batch of 50 instances, the allocation rate dropped to 20\%, indicating that single-node SPS is not a reliable metric for multi-node deployments.

Figure~\ref{fig:sps-score-distribution-one-fifty-nodes} further illustrates this discrepancy. Each instance type with a single-node SPS of 3 was requested 50 times. The fraction of types achieving successful allocation decreases sharply as the required number of nodes in the horizontal axis grows. Fewer than half of the types succeed when 10 or more instances are requested, and no type achieved full allocation at the maximum request of 50 instances.

This behavior arises because instances of the same type within an Availability Zone (AZ) are provisioned from a shared capacity pool~\cite{aws-spot-best-practices}. The single-node SPS reflects only whether at least one instance can be allocated from this pool, and does not capture how many instances the pool can simultaneously satisfy. As a result, a high single-node SPS does not guarantee the availability of a large multi-node request, since the request may exceed the remaining capacity of the shared pool. This gap between single-node and multi-node availability cannot be inferred from single-node SPS alone, motivating the need for a dataset that directly characterizes multi-node availability. 

\section{Collecting a Batch of Spot Instances Availability Dataset}\label{sec:collecting-batch-availability-dataset}

AWS offers SPS values for multiple instances of up to 50, but its query limitations make it difficult to retrieve this data efficiently. For example, within a 24-hour window, only 50 distinct query scenarios are allowed, and queries for the same configuration with different node counts are treated as separate requests. This restriction results in significant query overhead when gathering data for all combinations of instance types, regions, and node counts. Even using the optimal query distribution proposed by SpotLake~\cite{spotlake-iiswc}, approximately 3,300 queries are needed for all instance types for a single-node with 66 separate accounts. To query more than a single-node, the query overhead and required accounts increase proportionally. For example, querying SPS values from one to 50 nodes would require 165,000 queries, spread across 3,300 accounts. Moreover, to maintain dataset timely, these queries must be run periodically, making it nearly impossible to query all possible combinations. To address this, we propose a sampling heuristic to query the dataset in a timely manner.

\subsection{Uniform Spacing Query Sampling}\label{sec:usqs}
To mitigate the high query overhead required to track multi-node SPS values, we propose a heuristic named \emph{Uniform Spacing Query Sampling} (USQS). Instead of querying the entire range of node counts in every collection cycle, USQS probes a single target node count, $T_c$, at each periodic interval of $p$ minutes. The target count is systematically alternated across cycles, incremented by a fixed step size, $T_s$.

Specifically, after querying for $T_c$ nodes, the subsequent query targets $T_c + T_s$ nodes. This process continues until the target count exceeds the predefined maximum, $T_{max}$, at which point it resets to the minimum, $T_{min}$. This method allows for granular control over query costs by adjusting both the query interval $p$ and the step size $T_s$. However, this approach introduces a trade-off where a specific node count is re-evaluated only after a delay of $(\lfloor \frac{T_{max} - T_{min}}{T_s} \rfloor + 1) \times p$ minutes. During this period, any SPS fluctuations could lead to data staleness, necessitating an analysis of this proposed method's data integrity.

\subsubsection{Assessing Sampled Dataset Integrity}
A key parameter in USQS is the step size, $T_s$, which determines the granularity of the sampling. A larger $T_s$ reduces query overhead but increases the risk of missing the exact transition points. Considering both query overhead and the correctness of the dataset, a step size of $T_s=5$ is adopted, which provides a favorable balance between capturing significant availability trends and maintaining low query overhead.

To quantitatively assess the potential information loss from this sampling strategy with the given $T_s$, entropy, a measure of uncertainty from information theory, is employed~\cite{entropy-information}. The entropy $H(X)$ is defined as

\begin{equation}
H(X) := -\sum_{x \in \mathcal{X}} p(x) \log p(x)
\end{equation}

\noindent where $X$ is a discrete random variable representing the SPS value observed at a given query point, $\mathcal{X}$ denotes the set of all possible SPS outcomes (i.e., $\{1, 2, 3\}$), and $p(x)$ is the probability of observing SPS value $x$. Intuitively, entropy quantifies the degree of unpredictability in SPS transitions: when all outcomes are equally likely, entropy reaches its maximum, indicating maximum uncertainty; conversely, when the distribution is heavily concentrated on a single outcome, entropy approaches zero, reflecting highly predictable behavior.

A higher entropy value signifies greater randomness in the data, which increases the risk of missing critical changes when sampling periodically. Conversely, data with lower entropy exhibits more predictable patterns, making it well-suited for a sampling-based collection approach. With $T_s=5$ and $T_{max}=50$, there are 11 possible discrete numbers of nodes, $\{1, 5, 10, \dots, 50\}$. If the distribution of these outcomes were uniform, the probability of each would be $\frac{1}{11}$, yielding a maximum possible entropy of 3.4594 bits. However, the real SPS value exhibits non-uniform, skewed distributions~\cite{spotlake-iiswc}.

To measure the actual entropy, multi-node SPS data was collected from 844 instance types between January 26th and 29th, 2025, and the collected dataset yielded a measured entropy of 2.5052 bits. This value is significantly lower than the theoretical maximum for a uniform distribution. The lower entropy confirms that the temporal behavior of score change patterns is not random but contains predictable patterns. This result validates that the USQS method can effectively capture the state of multi-node availability with minimal information loss while substantially reducing query overhead, a claim further substantiated in the evaluation section.

\subsection{Tracking Score Transition Point}\label{sec:tstp}
While the USQS method effectively reduces query overhead, its sampling nature implies it may not identify the exact node count at which an SPS score transition occurs. For scenarios where higher precision is critical, this section introduces an alternative approach, Tracking Score Transition Points (TSTP), designed to locate the precise transition points without the potential for sampling-induced data loss.

This method is based on the observation that the SPS for a given instance type within an AZ exhibits a monotonically non-increasing property. As the number of requested nodes increases, the SPS value either remains the same or decreases. The TSTP approach leverages this behavior to efficiently and accurately pinpoint the specific node counts where the SPS value changes from 3 to 2, or 2 to 1.

For a given range of node counts [$T_{min}$, $T_{max}$], two key transition points are defined:
\begin{itemize}[noitemsep, topsep=0em, labelsep=0.5em]
    \item $T3$: The largest node count for which the SPS is 3.
    \item $T2$: The largest node count for which the SPS is 2.
\end{itemize}
By definition, $T3 \leq T2$. These two values concisely represent the multi-node availability metric; any request for a number of nodes up to $T3$ will have an SPS of 3, while requests between $T3+1$ and $T2$ will have an SPS of 2, and so on. A standard binary search algorithm can identify these points within $O(\log(T_{max} - T_{min}))$ API queries, avoiding the need to query every possible node count.

\paragraph{Reducing Query Overhead via Caching and Early Stopping}
To reduce the query overhead of a standard binary search, two optimization techniques are applied. First, since SPS values for an instance type tend not to fluctuate drastically over short periods~\cite{spotlake-iiswc}, the $T3$ and $T2$ values from the previous data collection cycle are cached. In the next cycle, the search begins near the cached value rather than at the midpoint of the entire [$T_{min}$, $T_{max}$] range, allowing the algorithm to narrow the search range with a single API call. Second, an \emph{early stopping} mechanism terminates the binary search when the search range ($T_{high} - T_{low}$) becomes smaller than a predefined threshold~$e$, as an approximate transition point within a small error margin is sufficient for assessing instance stability. These two optimizations are complementary: caching accelerates the early stages of the search by leveraging temporal continuity, while early stopping eliminates the diminishing-return queries in the final stages where additional query yields only a marginal approximate error reduction.

\section{Recommendation Engine for Multi-Node Spot Instances}
While real-time availability datasets from cloud vendors assist in the initial selection of spot instances, this information alone is insufficient for identifying configurations that are both cost-efficient and reliable over time. Raw availability data does not capture historical stability patterns or provide a comparative cost analysis across diverse instance types. To address this gap, this section details a recommendation engine that systematically identifies optimal multi-node spot instance pools by leveraging historical multi-node availability and price datasets.

The recommendation process begins with user-specified requirements for a computing resource pool, primarily the total number of CPU cores ($R_C$) or the total memory size ($R_M$). Users can also apply optional filters to narrow the set of candidate instances based on criteria such as instance family, category, or region. Each candidate node is then evaluated using a cost score and an availability score, which serve as the basis for recommendations.

\subsection{Quantifying Spot Instance Cost}
To establish a quantitative metric for comparing spot instance costs, the total cost to fulfill a user's request with a specific instance type must first be calculated. Let $p_i$ be the unit price for an instance type $i$ with $CPU_i$ cores. For a resource request defined by a total number of CPU cores, $R_C$, the required number of instances, $N_i$, is calculated as $N_i = \lceil R_C / CPU_i \rceil$. Consequently, the total cost for a pool composed of instance type $i$ is $c_i = p_i \times N_i$.

Let $\mathcal{C}$ be the set of total costs for all candidate instances. To enable a fair comparison, these costs are normalized to a score, $CS_i$, on a scale of [0.0, 100.0] using Equation~\ref{eq:cost-score}.

\begin{equation}
CS_i = 100\times \frac{C_{min}}{C_i}
\label{eq:cost-score}
\end{equation}

In the equation, $C_{min}$ denotes the minimum of the cost set $\mathcal{C}$. The scoring function computes the ratio of the minimum cost to the cost of instance $i$, producing a score of 100 for the cheapest instance and proportionally lower scores for more expensive ones. The score directly reflects how cost-efficient an instance is relative to the best available option, without requiring any distributional assumptions.

A key advantage of this inverse min-scaling formulation is its independence from the overall cost distribution. Unlike MinMax scaling~\cite{scikit-learn-minmax-scaling}, which compresses scores when extreme outliers exist and exaggerates marginal differences when all costs are close, the proposed metric evaluates each instance solely against the minimum cost. This provides stable and interpretable scores regardless of the cost distribution's shape.

\subsection{Quantifying Spot Instance Availability}
To quantitatively assess the availability of a spot instance, a composite scoring model is proposed. This model is based on the time-series analysis of \textbf{$T3$}, defined as the maximum number of nodes for which the SPS score remains at 3. The availability score, $AS_i$, for a given instance type $i$ is derived from three key characteristics of its $T3_i$ time-series data including magnitude, trend, and volatility.

First, the overall \emph{magnitude} of availability is captured by the area under the $T3_i$ time-series curve over the observation period, normalized to a scale of [0.0, 1.0], denoted $A3_i$, using a MinMax scaler across all candidate instances. A higher $A3_i$ indicates a greater and more consistent capacity for multi-node deployments.

Second, the \emph{trend} of availability over time, denoted by $m_i$, provides a predictive measure of future stability. An increasing $T3$ trend is a positive indicator, while a decreasing trend signals potential risk. The trend is calculated as the normalized slope of a first-order linear regression model fitted to the $T3_i$ data~\cite{first-order-regression}.

Third, the \emph{volatility} of availability, represented by $\sigma_i$, quantifies the stability of the instance. High fluctuation of $T3$ is undesirable for reliable workloads. This metric is calculated as the normalized standard deviation of the $T3_i$ values. A larger $\sigma_i$ corresponds to greater instability.

These three components are combined to compute the final availability score, $AS_i$, as shown in Equation~\ref{eq:sps-availability-score}.

\begin{equation}\label{eq:sps-availability-score}
AS_i = 100 \times (A3_i \times (1.0 + \lambda \times (m_i - \sigma_i)))
\end{equation}

The formula begins with the base magnitude score $A3_i$ and adjusts it through a scaling coefficient $\lambda$ that bounds the maximum influence of the trend and volatility components. Both $m_i$ and $\sigma_i$ are normalized to $[0.0, 1.0]$, so the adjustment term $(m_i - \sigma_i)$ is bounded within $[-1.0, 1.0]$. The coefficient $\lambda$ therefore determines the maximum percentage by which the base score can be adjusted. Extensive empirical evaluation reveals that $\lambda$ being 0.1 shows the best performance for spot instance reliability modeling, under which the formula applies a bonus of up to 10\% for a positive trend and a penalty of up to 10\% for high volatility. The justification for this choice is presented in the evaluation section.

\begin{figure}[t]
    \centering
    \subfloat[consistently high \textbf{$T3$} (score : 100)]{
        \includegraphics[width=0.235 \textwidth]{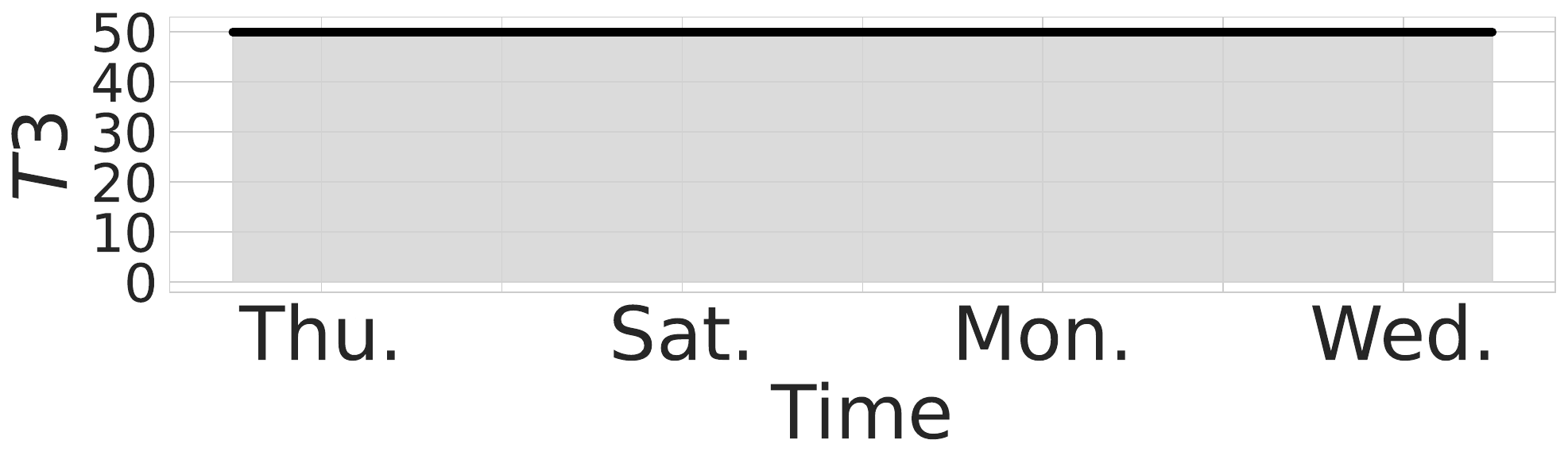}
        \label{fig:availability-score-explanation-high-t3}
    }
    \subfloat[consistently low \textbf{$T3$} (score : 0)]{
        \includegraphics[width=0.235 \textwidth]{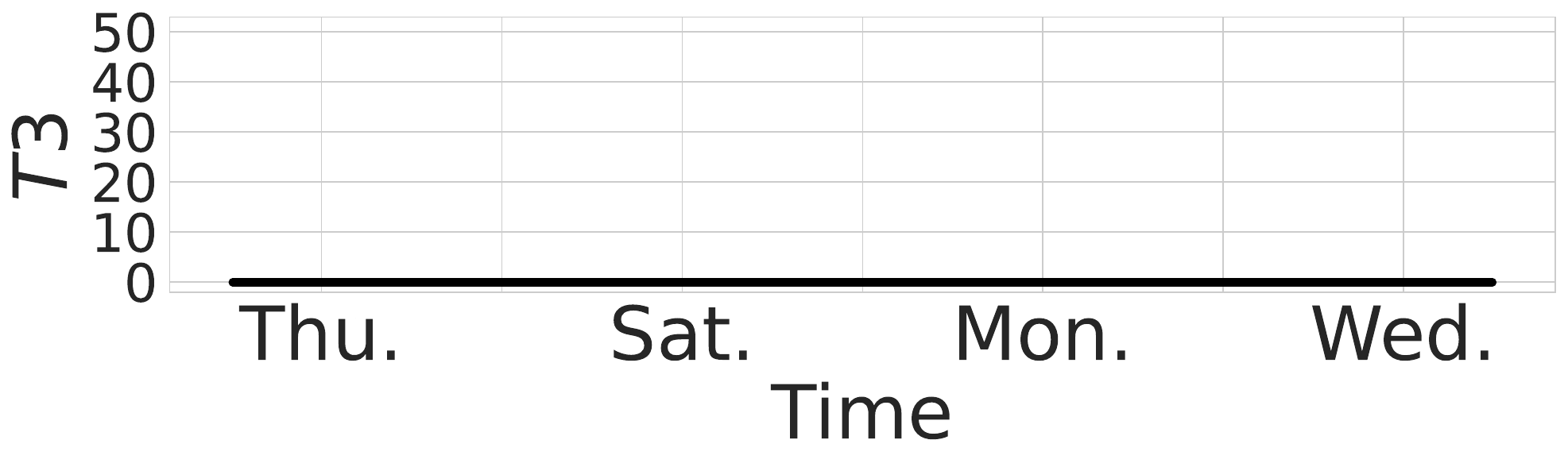}
        \label{fig:availability-score-explanation-low-t3}
    }\\
    \subfloat[positive slope \textbf{$T3$} (score : 59)]{
        \includegraphics[width=0.235 \textwidth]{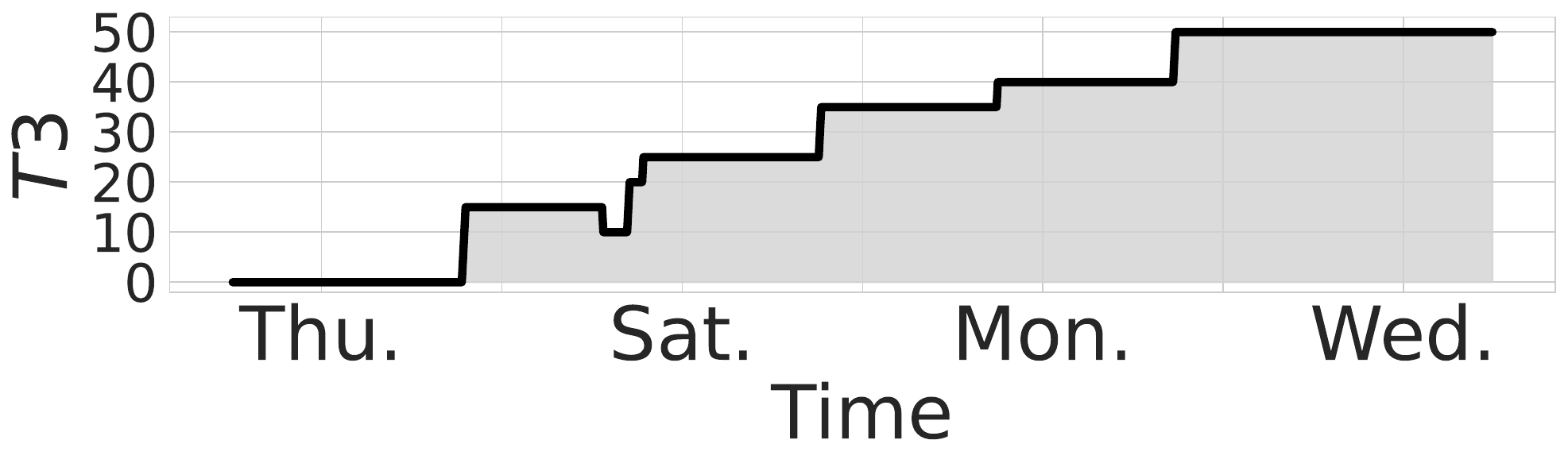}
        \label{fig:availability-score-explanation-upward-slope}
    }
    \subfloat[periodic changing \textbf{$T3$} (score : 45)]{
        \includegraphics[width=0.235 \textwidth]{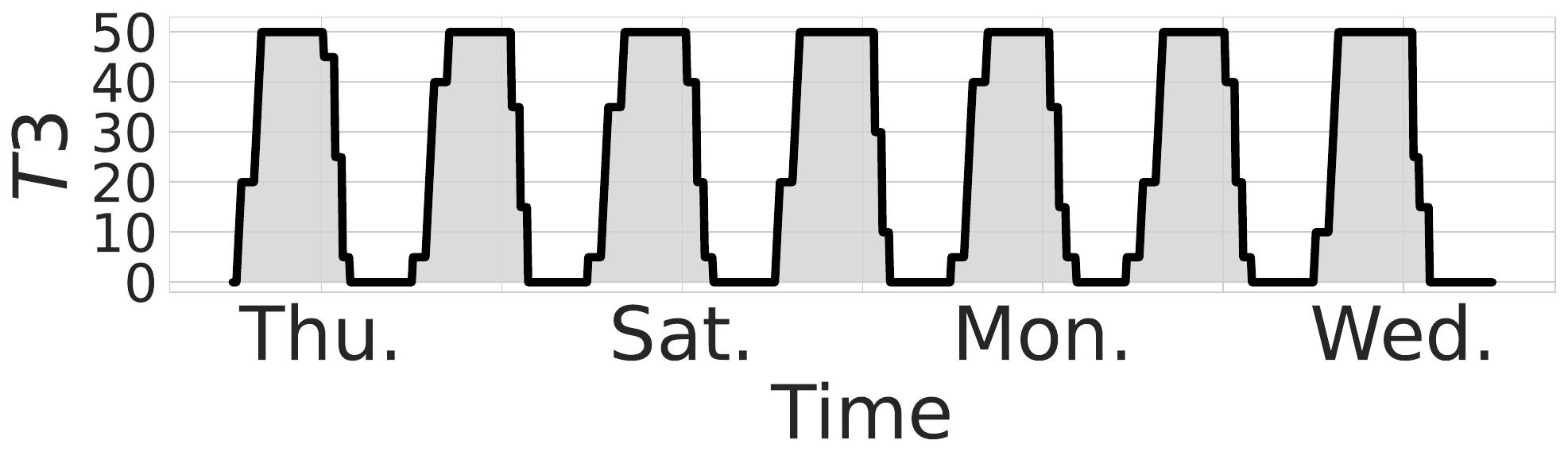}
        \label{fig:availability-score-explanation-regular-change}
    }
    \caption{Example of quantifying spot instance availability score by using the area, slope, and standard deviation of SPS score change over time}
    \label{fig:availability-score-explanation}
\end{figure}

Figure~\ref{fig:availability-score-explanation} provides examples of how the availability score ($AS_i$) quantifies different temporal patterns in $T3$ values. The horizontal axis in each plot represents time, and the vertical axis represents the $T3$ value, capped at a maximum of 50 in this scenario. An instance with a \emph{consistently high} $T3$ value (Figure~\ref{fig:availability-score-explanation-high-t3}) achieves a perfect score of 100. This is the result of a maximal area component ($A3_i$), combined with zero penalty for volatility ($\sigma_i=0$) and no adjustment for trend ($m_i=0$). Conversely, a consistently low $T3$ (Figure~\ref{fig:availability-score-explanation-low-t3}) yields a score of 0 because its base area score is zero.

Dynamic patterns are scored based on a combination of factors. The instance in Figure~\ref{fig:availability-score-explanation-upward-slope} displays a positive trend. Although it incurs a minor penalty for its volatility ($\sigma_i > 0$), it receives a substantial bonus for the upward slope ($m_i > 0$), resulting in a favorable score. In contrast, the instance with periodic fluctuations (Figure~\ref{fig:availability-score-explanation-regular-change}) is heavily penalized for its high volatility. With a slope that converges to zero ($m_i \approx 0$), it receives no trend-based bonus, leading to a comparatively low score. These examples demonstrate how the $AS_i$ score holistically evaluates availability by integrating the magnitude, trend, and stability of an instance's $T3$ profile.

\subsection{Spot Instance Recommendation}\label{sec:recommendation-engine}
The recommendation engine first calculates the availability score ($AS_i$) and the cost score ($CS_i$) for all candidate instances that meet the user's initial requirements. These scores are then combined into a final score, $S_i$, using a tunable weight parameter, $W \in [0.0, 1.0]$, as shown in Equation~\ref{eq:total-score}.
\begin{equation}\label{eq:total-score}
    S_i= W \times AS_i + (1.0 - W) \times CS_i 
\end{equation}
The weight $W$ allows users to prioritize availability over cost, or vice versa. Empirical analysis, detailed in the evaluation section, indicates that a weight of $W=0.5$ provides a robust balance between the two objectives, and it is used as the system's default setting.

Simply selecting the single instance type with the highest score ($S_i$) is often suboptimal in a multiple spot instances resource pool. Fulfilling a large resource request with numerous instances of a single instance type makes the entire workload vulnerable to an interruption event due to the specific type. To mitigate this risk, the proposed system constructs a heterogeneous pool of recommended instance types.

Given the scores for all candidate instances, the goal is to select a subset of instance types and determine the number of nodes for each, forming a pool that maximizes the total quality score while satisfying the user's resource requirements. This problem can be formulated as an Integer Linear Programming (ILP) problem, which is known to be NP-hard~\cite{ilp-np-hard}. While ILP solvers can identify globally optimal solutions for the score maximization objective, this formulation alone does not address the need for instance type diversity. Relying on a small number of high-scoring types makes the resulting pool vulnerable to correlated interruption events, which motivates the explicit consideration of diversity during pool construction.

Incorporating diversity into the ILP formulation, however, is structurally difficult. The number of selected types is not a linear function of the allocation variables, and expressing this requirement requires additional binary indicator variables with linking constraints. The resulting search space grows rapidly as the number of candidate instances increases, which makes ILP-based formulations impractical for real-time recommendation in large candidate spaces, where a quantitative comparison is presented in Section~\ref{sec:evaluations}.

\begin{algorithm}[ht]
\caption{Greedy Heuristic for Spot Instance Pool Formation}
\label{alg:pool-formation}
\begin{algorithmic}[1]
\Procedure{FormHeterogeneousPool}{$C, R_{req}$}
    \State \textbf{Input:} $C$: Set of candidate instances with scores $S_i$.
    \State \textbf{Input:} $R_{req}$: Total required resources (e.g., CPU cores).
    \State \textbf{Output:} $X_{best}$: A map of instance types and their count.
    
    \State $C_{sorted} \gets \text{Sort } C \text{ by score } S_i \text{ in descending order}$
    \State $P \gets \emptyset$ \Comment{The set of instance types in the current pool}
    \State $X_{best} \gets \emptyset$
    \State $x_{prev\_top} \gets \infty$ \Comment{Allocation of the top-ranked instance}

    \For{$i \in C_{sorted}$}
        \State $P \gets P \cup \{i\}$ \Comment{Add next best instance to the pool}
        \State $S_{total} \gets \sum_{j \in P} S_j$
        \State $X_{curr} \gets \text{new map}$
        
        \ForAll{$j \in P$} 
            \State $R_j \gets \frac{S_j}{S_{total}} \times R_{req}$ \Comment{Score-based allocation}
            \State $x_j \gets \lceil \frac{R_j}{CPU_j} \rceil$
            \State $X_{curr}[j] \gets x_j$
        \EndFor
        
        \If{$X_{curr}[C_{sorted}[0]] \ge x_{prev\_top}$ \textbf{or} $X_{curr}[i] = 0$} 
            \State \textbf{return} $X_{best}$
        \EndIf
        
        \State $X_{best} \gets X_{curr}$ 
        \State $x_{prev\_top} \gets X_{curr}[C_{sorted}[0]]$
        
    \EndFor
    \State \textbf{return} $X_{best}$
\EndProcedure
\end{algorithmic}
\end{algorithm}

To address this issue, SpotVista introduces a greedy heuristic that enhances the diversity of the selected instance types while minimizing the degradation of the aggregated total scores. The detailed procedure is presented in Algorithm~\ref{alg:pool-formation}.

The algorithm begins by sorting all candidate instances in descending order of their final scores with initialization of variables (Lines 5-8), including the recommendation pool $P$ and the previous allocation for the top-ranked instance, $x_{prev\_top}$.

The core of the algorithm is a main loop that iterates through the sorted candidates (Line 9). In each iteration, the next highest-scoring instance type is added to the pool $P$ to increase diversity (Line 10). Subsequently, the system recalculates the node allocation for every instance currently in the pool (Lines 13-17). This is achieved by distributing the total required resources, $R_{req}$, among the pool members proportionally to their individual scores (Line 14) and then calculating the necessary number of nodes for each (Line 15).

After each reallocation, the algorithm evaluates two termination conditions (Line~18). The first, $X_{curr}[C_{sorted}[0]] \ge x_{prev\_top}$, detects that the top-ranked instance's allocation has stopped decreasing, meaning the newly added instance's score is too low to meaningfully redistribute resources away from the dominant type, and further additions of even lower-scored instances would yield diminishing diversification. The second, $X_{curr}[i] = 0$, indicates that the latest addition receives no allocation under the score-proportional distribution (Line~14), so it would contribute no capacity to the pool. When either condition is met, the algorithm returns the previous iteration's allocation, $X_{best}$ (Line~19), preserving the last state in which diversification was effective.

If the algorithm does not terminate, it updates $X_{best}$ with the current allocation and saves the new top-ranked node count to $x_{prev\_top}$ for the next iteration's comparison (Lines 21-22). This process ensures that the recommendation is iteratively refined to balance score quality with instance diversity.

\section{SpotVista Implementation}

\begin{figure}[t]
    \centering
    \includegraphics[width=0.98\columnwidth]{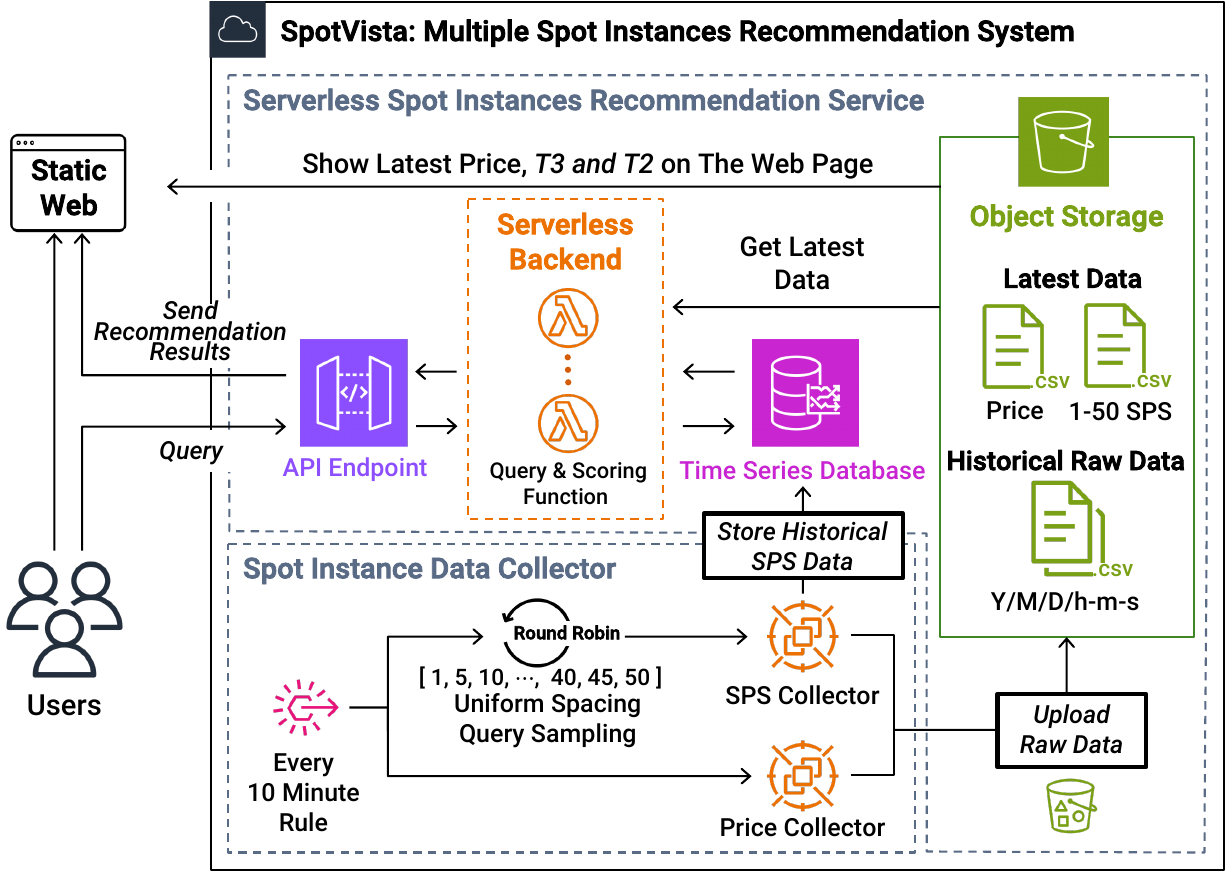}
    \caption{Implementation of the proposed system}
    \label{fig:SpotVista-architecture}
\end{figure}

SpotVista is publicly available as a web-service. The architecture of the implemented system targeting spot instances is shown in Figure~\ref{fig:SpotVista-architecture}. The \emph{Data Collector} component at the bottom applies the proposed USQS heuristic for SPS data collection where queries are periodically generated every $10$ minutes with the target number of nodes ranging from $5$ to $50$, incremented by $5$. For price data, we directly utilize the price dataset provided by cloud vendors. The collected dataset is stored in an object storage, available for direct access by users or for use by the recommendation module. The files are saved with a Year-Month-Date-Time naming convention to include historical data.

The recommendation module is designed using a serverless architecture~\cite{serverless-backward-forward} to flexibly handle irregular recommendation requests from users. When users provide their resource requirement, such as minimum number of CPU cores or memory size, along with optional information such as specific instance types, regions, or the maximum number of returned instance types, this data is sent to an API Endpoint service, which acts as the endpoint for the recommendation service. It forwards the request to a Function-as-a-Service platform, which filters the instance types that meet the user's requirements and retrieves the historical $T3$ values of the target instances via a serverless time-series database service. The recommendation score is calculated using this information and then returned to the user. The web service itself is served as static HTML files via object storage, and the instance recommendation functionality interacts with the backend RESTful API through the Fetch API.

The monthly infrastructure cost of operating SpotVista is approximately \$55 USD, comprising \$30 for an instance running the data collection pipeline, \$20 for the time-series database service, and \$5 for object storage. The recommendation module incurs no additional fixed cost, as it is deployed on a serverless architecture that charges only per invocation. This low operational overhead demonstrates the practicality of continuously maintaining a large-scale multi-node availability dataset as a public service.

\begin{figure*}[t]
    \centering
    \subfloat[Different query heuristics comparison ]{
        \includegraphics[width=0.32\textwidth]{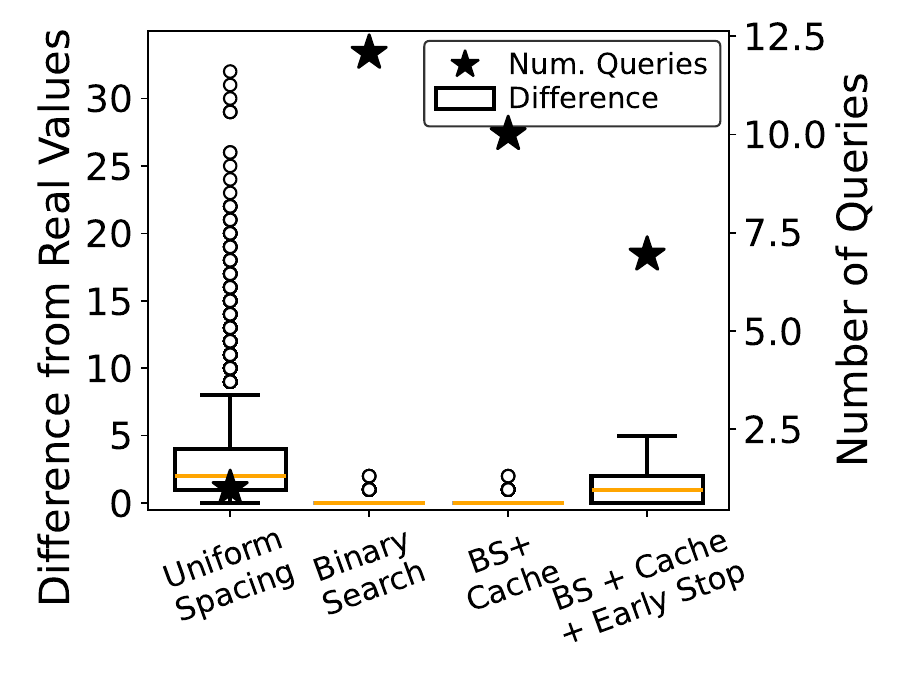}
    \label{fig:query-heuristic-comparison}
    }
    \subfloat[Overhead and error rate of USQS]{
        \includegraphics[width=0.32\textwidth]{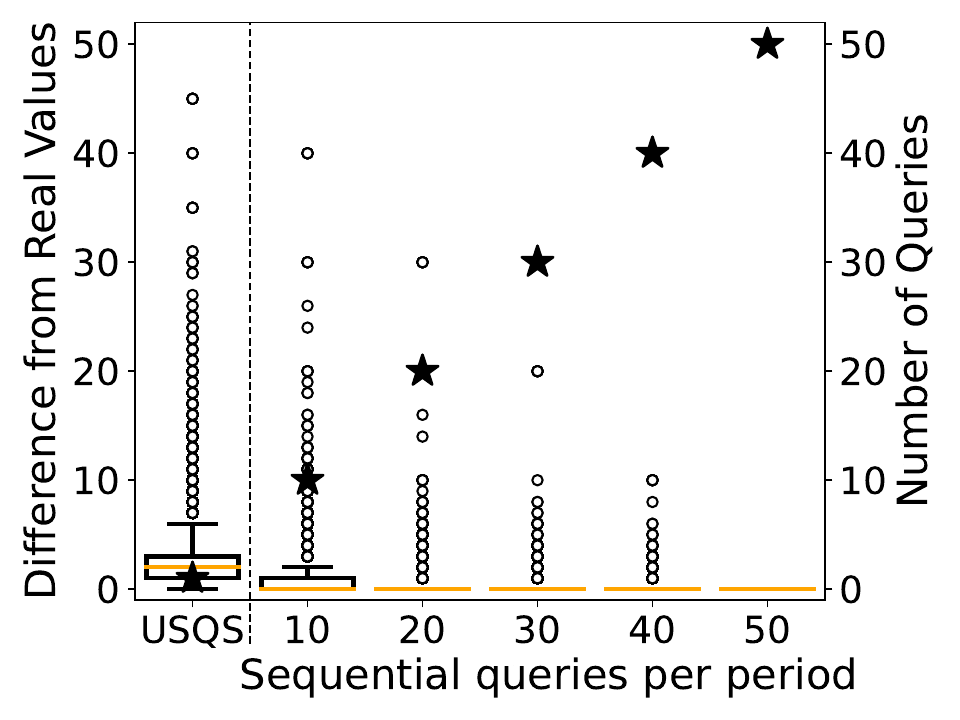}
    \label{fig:query-overhead-result-complete}
    }
    \subfloat[Error ratio analysis]{
        \includegraphics[width=0.32\textwidth]{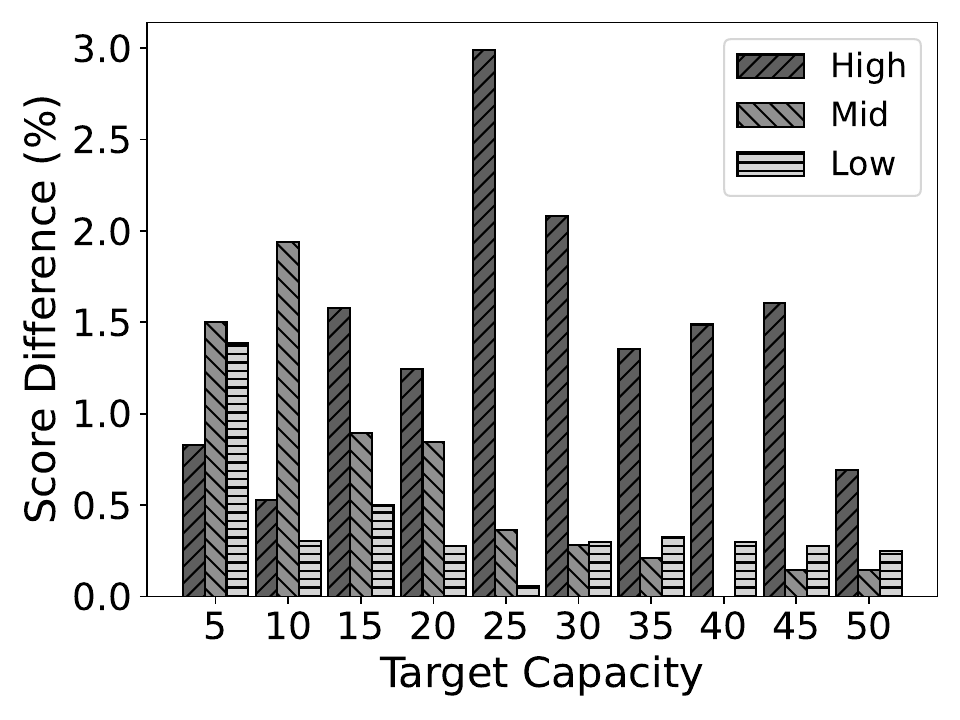}
    \label{fig:multi-sps-sampling-collection}
    }
    
    \caption{Multi-node SPS dataset collection overhead and completeness of the proposed sampling query heuristic}
    \label{fig:sampling-query-effectiveness}
\end{figure*}

\section{Evaluations}
\label{sec:evaluations}
We analyze the efficiency and operational overhead of the multi-node SPS data collection module. We also demonstrate the effectiveness of the instance recommendation algorithm through extensive experiments in real-world spot instance environments. An intuitive method to assess spot instance stability and interruption probability is to continuously run spot instances and record any interruptions~\cite{interrupt-visible-www, google-cloud-empirical-preemption, spot-instance-interrupt-check-cloud-2018}. Such an approach is valid for small-scale experiments. However, the cost can increase significantly as the scale of the experiment grows. To control the cost of interruption experiments, Wu et al.~\cite{cant-be-late} demonstrated that continuously running spot instances is not necessary. Instead, periodically sending spot instance requests and observing whether the requests succeed can effectively model spot instance stability. Li et al.~\cite{sky-nomad} adopted the same probing-based methodology to characterize regional spot availability. It issues lightweight launch requests that immediately terminate upon success, thereby measuring availability at low cost. Spot-and-Scoot~\cite{ddd-preprint} further corroborated this finding. They reported that spot request outcomes rarely overestimate actual capacity and that interruptions of co-located instances of the same type exhibit strong temporal correlation.

Given the large-scale experiments involving many instance types in this paper, we adopt this methodology. We periodically sent spot requests, recorded success or failure, and generated interruption data based on these observations. Using the result, we aim to answer the following research questions.

\begin{itemize}[noitemsep, topsep=0em, leftmargin=*, labelsep=0.5em]
    \item \textbf{RQ-1} Does the sampling-based multi-node SPS collection heuristic effectively capture key information without significant loss?
    \item \textbf{RQ-2} What characteristics are observed in the multi-node SPS dataset?
    \item \textbf{RQ-3} Does the proposed availability scoring mechanism work as expected?
    \item \textbf{RQ-4} How does the spot instance recommendation algorithm perform compared to the state-of-the-art approaches?
\end{itemize}

\subsection{Effectiveness of the SPS Query Heuristics}
\label{sec:effect-of-sps-query}
To answer \textbf{RQ-1}, this section evaluates the trade-off between query overhead and data integrity for the proposed heuristics. Figure~\ref{fig:sampling-query-effectiveness} presents this analysis using a dataset collected from October 12--15, 2024, for 15 instance types across 7 regions. The ground truth was established by a Full Scan that queries the SPS of all 50 target node counts every 10 minutes.

\paragraph{Analysis of Different Query Heuristics}
Figure~\ref{fig:query-heuristic-comparison} compares the $T3$ estimation error (primary vertical axis) and query overhead (secondary vertical axis) across four strategies in the horizontal axis. Plain Binary Search (BS) achieves minimal error but requires 12 queries per cycle. Adding caching and early stopping ($e{=}4$) reduces this to about 7 queries with a negligible average error of 0.9. USQS requires only a single query per cycle; its median error remains low, though the 100-minute re-query interval can cause larger deviations for highly dynamic instances. This marginal precision loss is a tolerable trade-off for the substantial reduction in query overhead.

\paragraph{Query Overhead vs.\ Error Rate}
Figure~\ref{fig:query-overhead-result-complete} compares USQS against sequential scanning with increasing query counts (10--50 per cycle). As the number of queries increases linearly, the $T3$ error decreases only marginally, confirming that the $10\times$--$50\times$ overhead reduction of USQS far outweighs its minimal precision loss.

\paragraph{Impact on Collected SPS Data}
Figure~\ref{fig:multi-sps-sampling-collection} shows the percentage difference in average SPS values between USQS and the Full Scan, categorized by historical SPS volatility (High, Mid, Low). Regardless of volatility, the maximum deviation remains below 3\%, confirming that the USQS sampling has a negligible effect on the integrity of the collected time-series data used by the availability scoring model.

\paragraph{Sensitivity to Step Size}
\begin{figure}[t]
    \centering
    \includegraphics[width=\columnwidth]{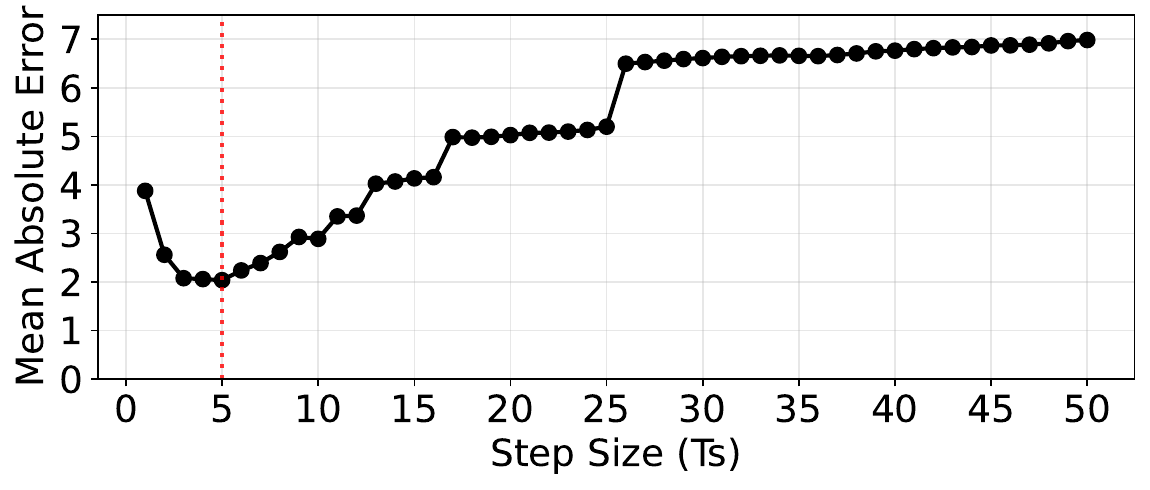}
    \caption{Mean Absolute Error of the USQS heuristic as a function of step size}
    \label{fig:usqs-step-size}
\end{figure}

\begin{figure*}[t]
    \centering
    \subfloat[$T3$ change patterns across different AWS region (2025)]{
        \includegraphics[width=\columnwidth]{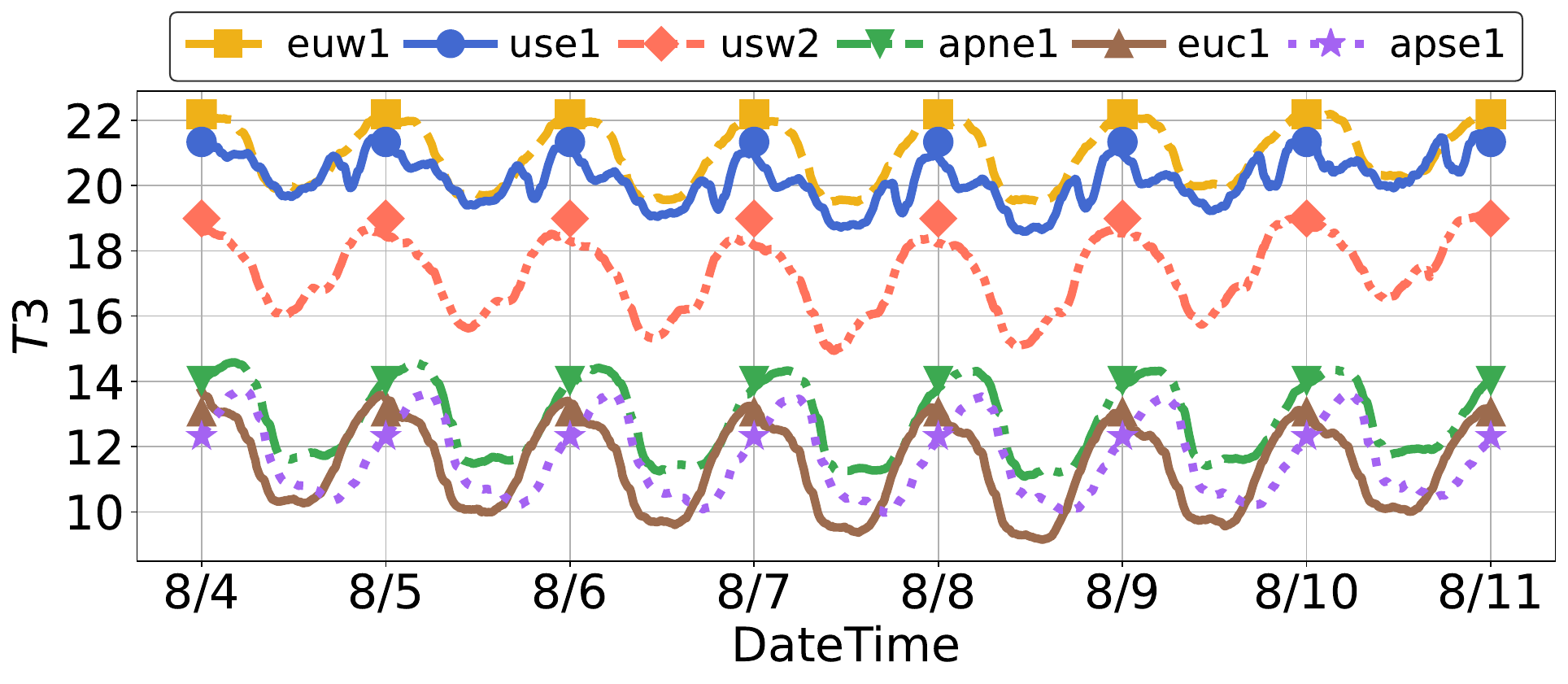}
        \label{fig:t3_change_pattern_region}
    }
    \subfloat[$T3$ change patterns by AZ in the AWS us-east-1 region (2025)]{
        \includegraphics[width=\columnwidth]{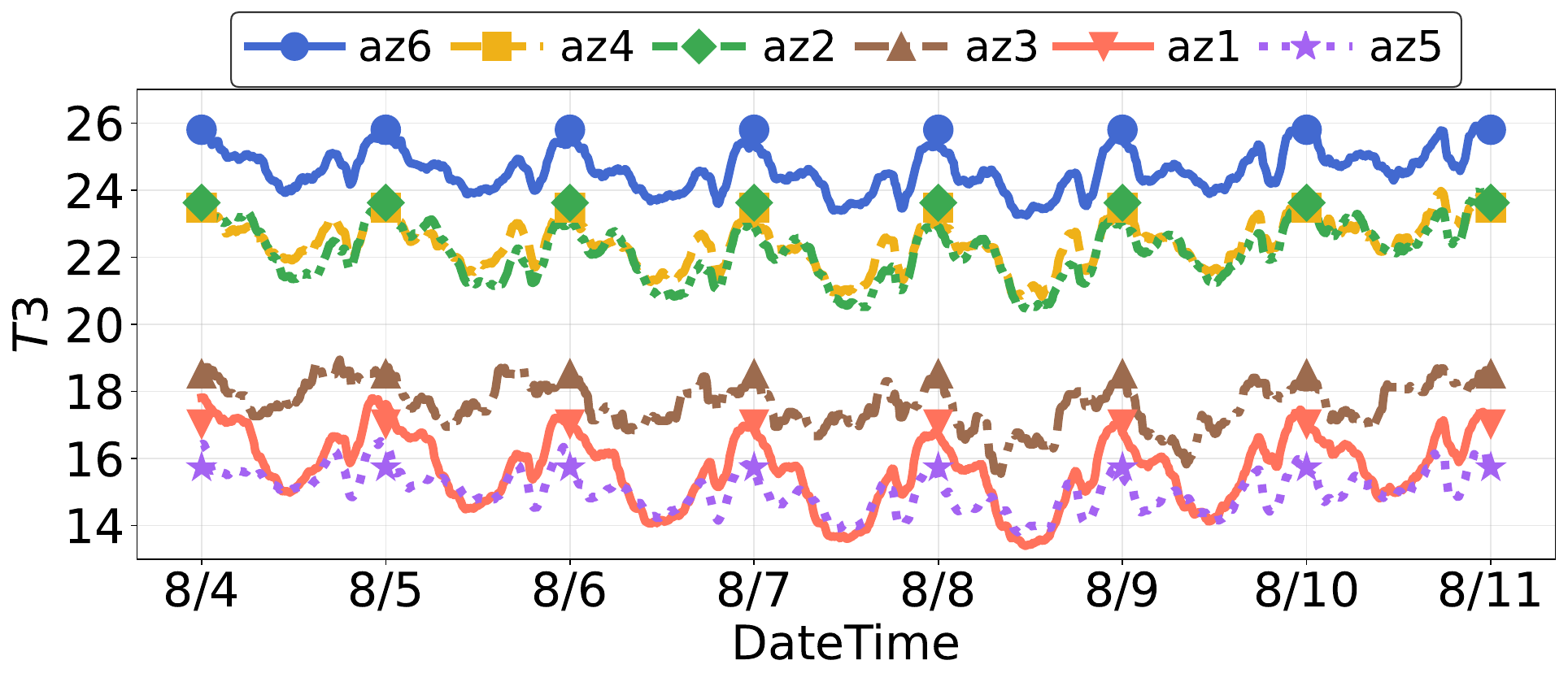}
        \label{fig:t3_change_pattern_az}
    }\\
    \subfloat[Weekly seasonal component of $T3$ extracted via MSTL decomposition for AWS and Azure]{
        \includegraphics[width=\textwidth]{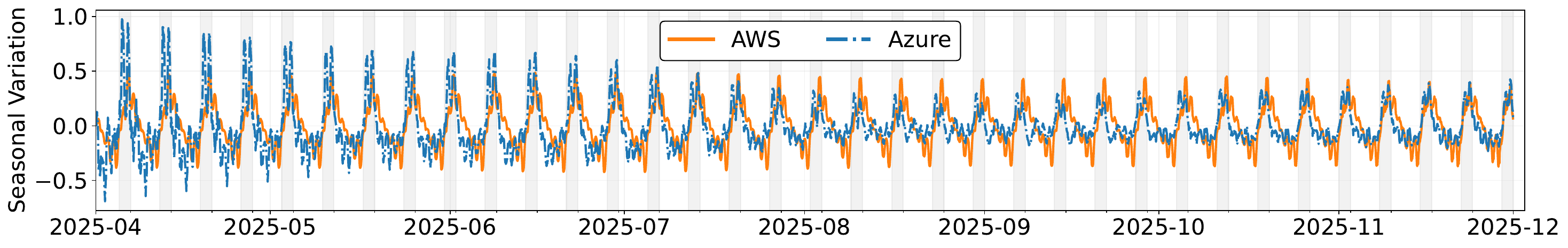}
        \label{fig:t3_change_mstl_seasonal}
    }
    \caption{Spatial, temporal, and seasonal characteristics of multi-node $T3$ availability, combining short-term observations with long-term MSTL decomposition~\cite{mstl}.}
    \label{fig:t3-region-az-pattern-sps}
\end{figure*}

To further validate the choice of $T_s = 5$, we simulated the USQS heuristic over the ground-truth dataset for all step sizes from $T_s = 1$ to $T_s = 50$. Figure~\ref{fig:usqs-step-size} shows that the MAE follows a U-shaped curve driven by two competing error sources: for small $T_s$, the prolonged round-robin cycle causes temporal staleness (e.g., 500 minutes per cycle at $T_s = 1$), while for large $T_s$, wide spacing between probes misses SPS transition points. The MAE reaches its minimum region at $T_s = 3$--$5$, where values remain below 2.1. Among these, $T_s = 5$ is selected because it requires the fewest probe points per cycle (11 points for $\{1, 5, 10, \ldots, 50\}$ versus 17 for $T_s = 3$), further reducing query overhead while maintaining near-optimal accuracy.

\begin{table}[t]
\caption{Temporal stability analysis of $T3$ seasonal patterns}
\label{tab:temporal-stability}
\centering
\small
\begin{tabularx}{\linewidth}{lXX}
\toprule
\textbf{Metric} & \textbf{AWS} & \textbf{Azure} \\
\midrule
\multicolumn{3}{l}{\textit{MSTL Variance Decomposition}} \\
\quad Daily (24h)  & 0.783 & 0.030 \\
\quad Weekly (168h) & 0.039 & 0.041 \\
\quad Trend        & 0.060 & 1.115 \\
\quad Residual     & 0.003 & 0.029 \\
\midrule
\multicolumn{3}{l}{\textit{Seasonal Strength ($F_S$)}} \\
\quad Daily        & 0.997 & 0.510 \\
\quad Weekly       & 0.934 & 0.608 \\
\midrule
\multicolumn{3}{l}{\textit{Bai-Perron Amplitude Stability}} \\
\quad Daily breakpoints   & 4          & 5          \\
\quad Daily max variation & $\pm$9\%   & $\pm$44\%  \\
\quad Weekly breakpoints  & 3          & 4          \\
\quad Weekly max variation & $\pm$7\%  & $\pm$28\%  \\
\bottomrule
\end{tabularx}
\end{table}

\subsection{Multiple Spot Instances Dataset Analysis}
\label{sec:multi-node-dataset-analysis}
To address \textbf{RQ-2}, this section analyzes the spatial and temporal characteristics of the collected multi-node SPS
dataset across both short-term and long-term horizons, followed by further analyses on instance size correlation, resource pool diversity, and $T3$ characteristics.

\paragraph{Short-term Spatial and Temporal Characteristics}
\label{sec:short-term-analysis}
The dataset covers July 1, 2024 to August 31, 2025, with 952 unique instance types across 17 AWS regions. Figure~\ref{fig:t3_change_pattern_region} and~\ref{fig:t3_change_pattern_az} show the average $T3$ values during the week (August 4--11, 2025). A consistent daily cyclical pattern appears across all major regions, where $T3$ values are higher during local nighttime and lower during business hours, consistent with prior spot availability studies~\cite{snape-azure-spot-mixture, interrupt-visible-www}. The same pattern holds at the AZ-level within us-east-1. Both levels also reveal that certain locations consistently offer higher multi-node availability, confirming that location selection is a critical factor for reliable spot deployments.

\paragraph{Long-term Seasonal Stability}
\label{sec:long-term-analysis}
To verify whether these patterns persist over longer horizons and across vendors, MSTL decomposition~\cite{mstl} is applied to extended datasets from AWS (April--November 2025; 1,114 types, 17 regions) and Azure (same period; 1,705 types, 67 regions). Two metrics quantify the stability pattern; the seasonal strength $\mathcal{F}_S$~\cite{time-series-clustering}, which measures how strongly a periodic component dominates over residual noise ($[0.0,1.0]$), and the Bai-Perron structural break test~\cite{bai-perron}, which detects statistically significant shifts in seasonal amplitude.

Figure~\ref{fig:t3_change_mstl_seasonal} visualizes the extracted weekly seasonal component, and Table~\ref{tab:temporal-stability} summarizes the quantitative results. For AWS, the daily cycle dominates the overall variance, both daily and weekly $\mathcal{F}_S$ exceed 0.93, and the Bai-Perron test shows amplitude variation of at most $\pm 9\%$. These results confirm that the cyclical behavior observed in the short-term analysis persists over eight months, providing an empirical basis for the historical pattern-based availability scoring proposed in this work.

It is noticeable that Azure exhibits a structurally different pattern. The trend component dominates variance rather than seasonal cycles, $\mathcal{F}_S$ values are substantially lower (0.51 daily, 0.61 weekly), and amplitude shifts reach $\pm 44\%$. Directly applying the proposed scoring methodology to Azure would therefore require adaptation mechanisms to account for the weaker and less stable seasonality, which we further discuss in Section~\ref{sec:discussion}.

\begin{figure}[t]
  \centering
  \subfloat[The distribution of correlation value for the T3 and the instance sizes]{
    \includegraphics[width=0.465\columnwidth]{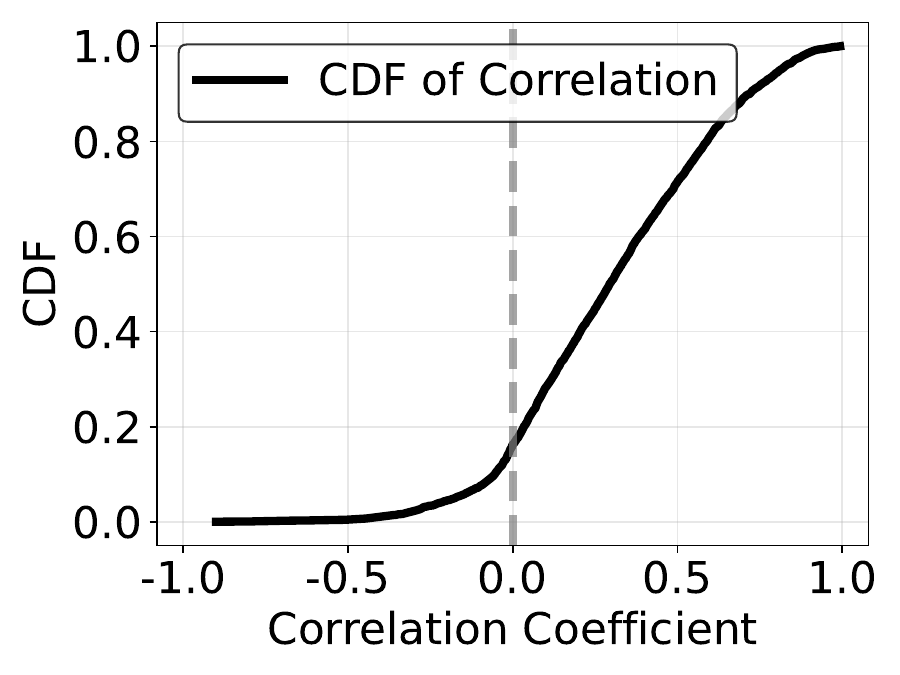}
    \label{fig:cdf-different-size-T3}
  }\hfill
  \subfloat[Proportion of time with higher T3 values across instance sizes]{
    \includegraphics[width=0.465\columnwidth]{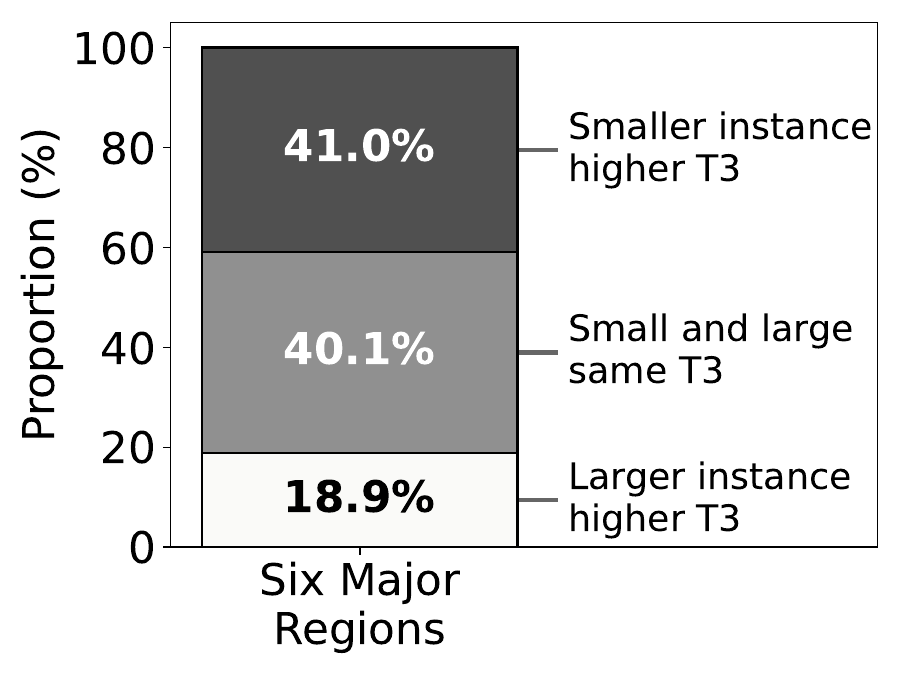}
    \label{fig:larger-instance-higher-T3}
  }
  \caption{Correlation of T3 values with respect to different instance sizes}
  \label{fig:sps-correlation-by-sizes}
\end{figure}

\begin{figure}[t]
  \centering
  \includegraphics[width=\columnwidth]{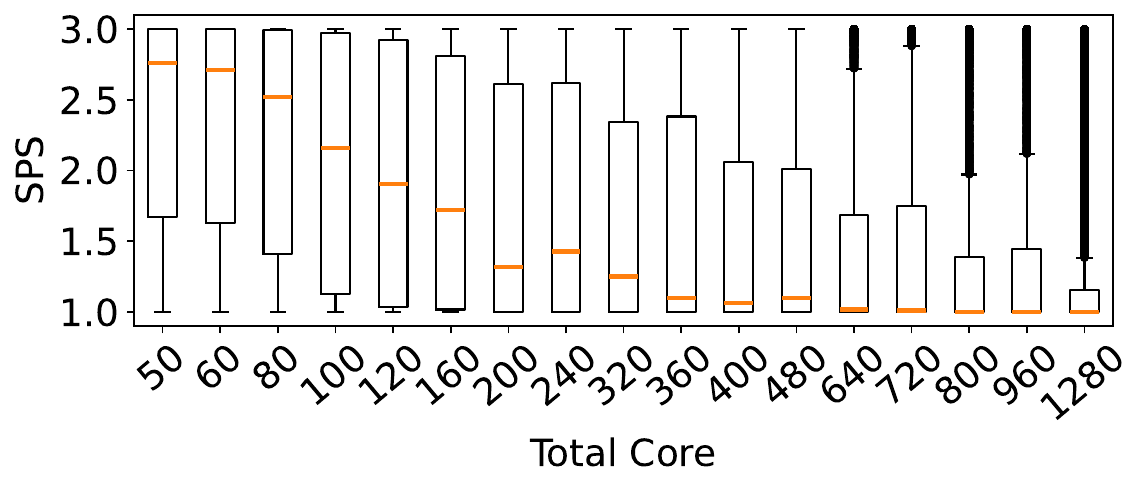}
  \caption{SPS value distribution diversity to build a compute resource pool}
  \label{fig:sps-difference-make-total-cores}
\end{figure}

\paragraph{Correlation of T3 Values and Instance Size}
To evaluate the correlation between the instance size and multi-node availability, $T3$ scores of 5,414 pairs of adjacent-sized instances from the same family (e.g., \texttt{m5.large} and \texttt{m5.2xlarge}) were compared using the August 2025 dataset. Figure~\ref{fig:cdf-different-size-T3} shows that 83.7\% of pairs exhibit a positive $T3$ correlation, indicating that availability patterns within the same family are generally synchronized across sizes; when the availability of smaller instances increases, the availability of larger instances tends to increase as well.

However, synchronized patterns do not imply equal availability. Figure~\ref{fig:larger-instance-higher-T3} shows that smaller instances had a higher $T3$ value 41.0\% of the time, while larger instances were superior only 18.9\% of the time, with the remaining 40.1\% being identical. Although smaller instances tend to offer better multi-node availability, larger instances still match or exceed them 59\% of the time. Considering that using a larger instance type requires fewer nodes to meet a compute demand, instance size alone is not a reliable predictor, underscoring the need to reference per-instance-type $T3$ rather than relying on size-based heuristics.

\noindent
\begin{figure}[t]
\begin{minipage}[t]{.485\linewidth}
  \centering
  \includegraphics[width=\linewidth]{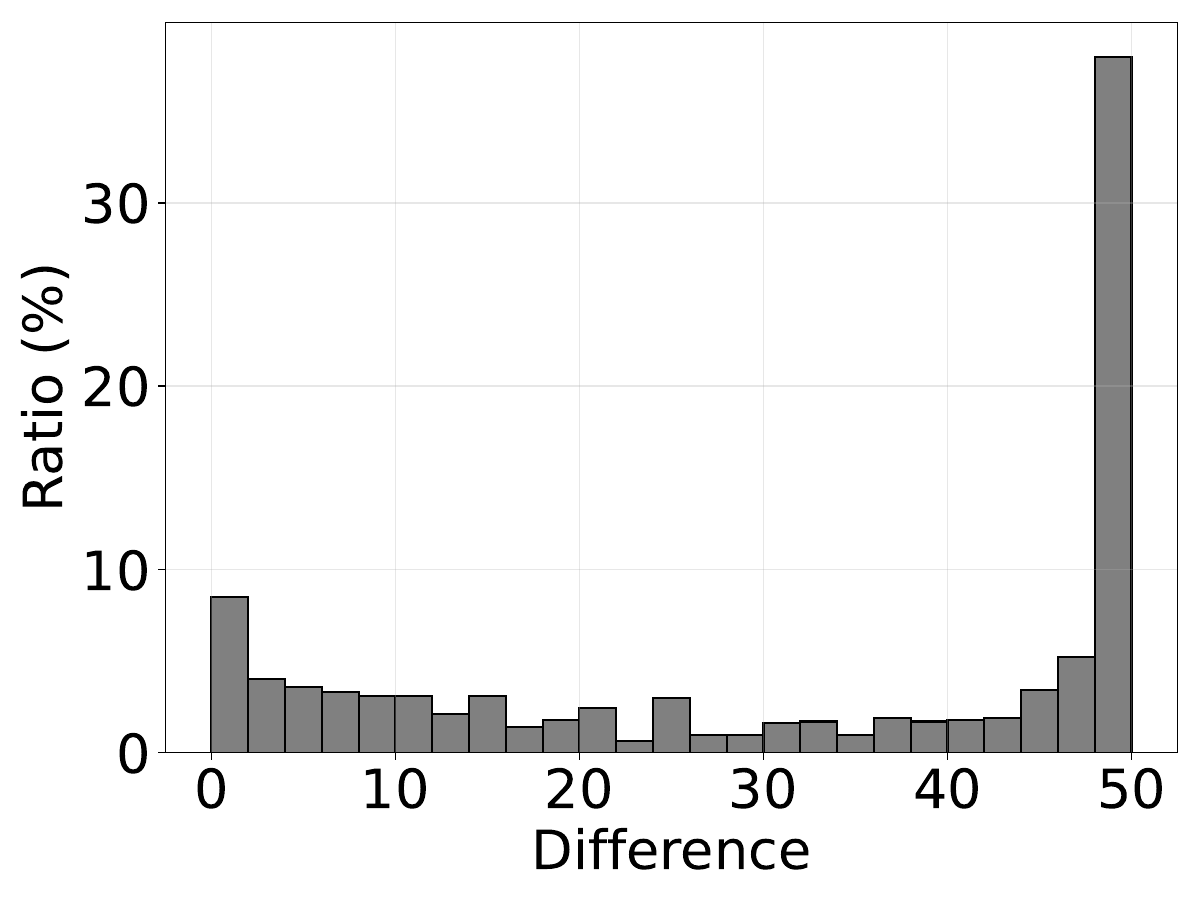}
  \captionof{figure}{T3 differences for the same instance type across regions and AZs}
  \label{fig:sps-3-max-instance-azs}
\end{minipage}\hfill
\begin{minipage}[t]{.485\linewidth}
  \centering
  \includegraphics[width=\linewidth]{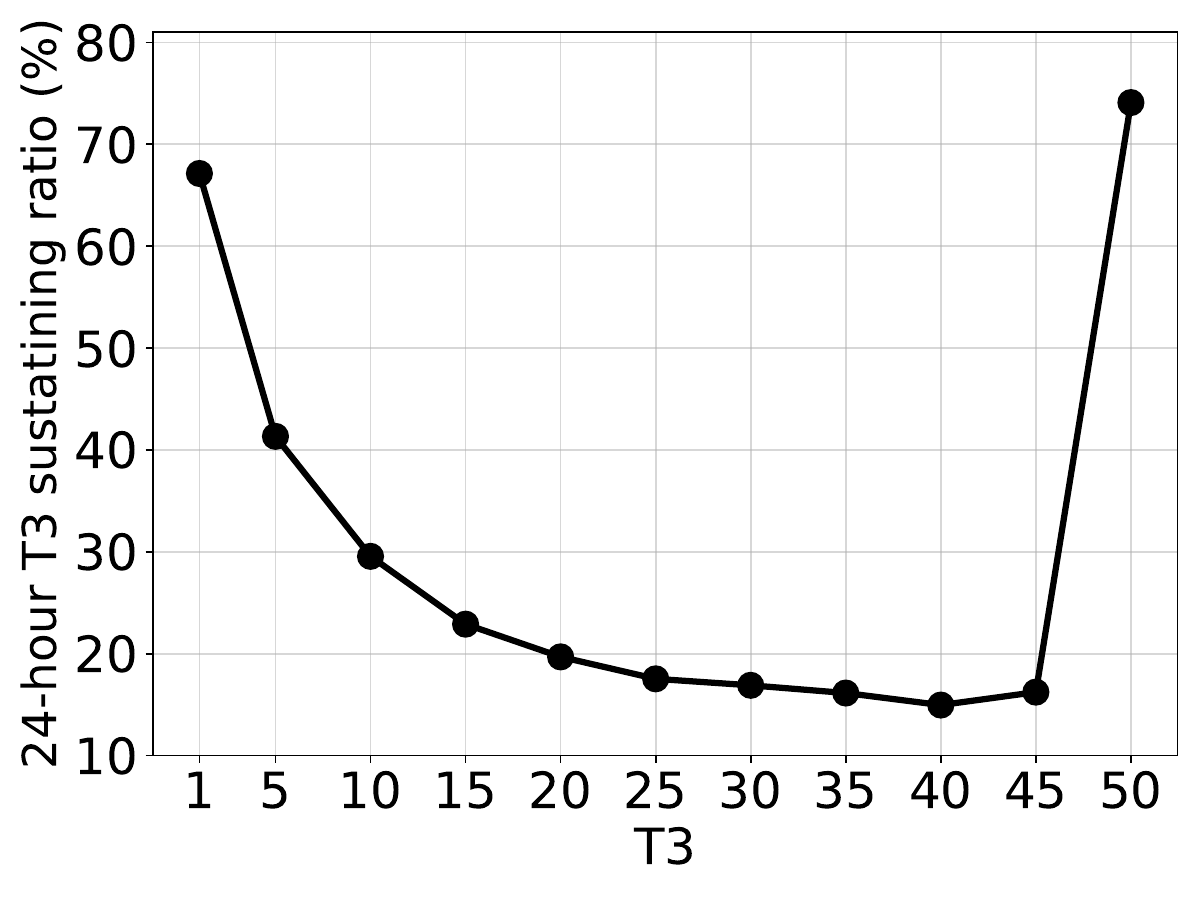}
  \captionof{figure}{The ratio of sustaining T3 values after 24 hours}
  \label{fig:24-hour-survive-T3}
\end{minipage}
\end{figure}

\paragraph{Diverse Spot Instance Availability}
Rather than pre-selecting a specific instance type, a more effective strategy is to define a total resource requirement
(e.g., target CPU cores) and explore the diverse instance combinations that can fulfill it. Figure~\ref{fig:sps-difference-make-total-cores} plots the SPS distribution for various instance combinations capable of fulfilling different total core count requirements. The median SPS shows an overall downward trend as the total number of required cores grows, falling below 2.0 around 120 cores and approaching 1.0 beyond 320 cores. Crucially, even for large requests, the upper quartiles and outliers reveal that high-SPS combinations still exist, though they might be harder to discover among the many low-availability options. This motivates the need for a recommendation engine that can identify these stable combinations from the multi-node SPS data.

\paragraph{Characteristics of T3}
Figure~\ref{fig:sps-3-max-instance-azs} examines spatial variation of multi-node availability for each instance type by calculating the difference between its maximum and minimum $T3$ values across all supported AZs whose value is shown in the horizontal axis. While some instance types show little variation, over 36\% exhibit the maximum possible difference of 50, which means that they have at least one AZ with $T3=50$ and another with $T3 \approx 0$. The choice of region and AZ is therefore a critical determinant to enhance multi-node stability.

Figure~\ref{fig:24-hour-survive-T3} plots the proportion of instances that sustain their initial $T3$ value over 24 hours, using 33,713 instance-type observations on August 1, 2025. The result follows a J-shaped curve, where for $T3$ values between 1 and 45, higher initial values are less likely to be sustained (e.g., 70\% sustaining ratio at $T3{=}1$ versus only 15\% at $T3{=}40$), suggesting that moderately high availability is often transient. This trend reverses sharply at $T3{=}50$, where the sustaining ratio reaches 74.1\%. This anomaly is likely a ceiling effect of the 50-node query limit where instances whose true capacity far exceeds 50 are capped at $T3{=}50$ and remain there even if their actual availability fluctuates above the threshold. While this is a limitation of the dataset, it also implies that a measured $T3$ of 50 is a signal of high stability.

\subsection{Effectiveness of Availability Scoring Mechanism}
This section addresses \textbf{RQ-3} by validating the effectiveness of the proposed availability scoring mechanism.

\paragraph{Experimental Setup} To measure real-world spot instance stability, experiments were conducted using 100 distinct instance types chosen to represent a wide range of predicted availability scores. From September 13 to October 9, 2024, 50 spot instances for each type were requested every 10 minutes, 24 hours a day. From the resulting spot instance request logs, a ground-truth metric, termed the \emph{Real Availability Score}, was calculated following the methodology proposed by Wu et al.~\cite{cant-be-late}. This real-world score was then compared against two predictors:
\begin{enumerate}[noitemsep, topsep=0em]
    \item The proposed composite \emph{Predicted Availability Score}.
    \item A baseline using the raw, single-point \emph{vanilla $T3$ value}.
\end{enumerate}

\begin{figure}[t]
    \centering
    \subfloat[Proposed availability score]{
    \includegraphics[width=0.47\columnwidth]{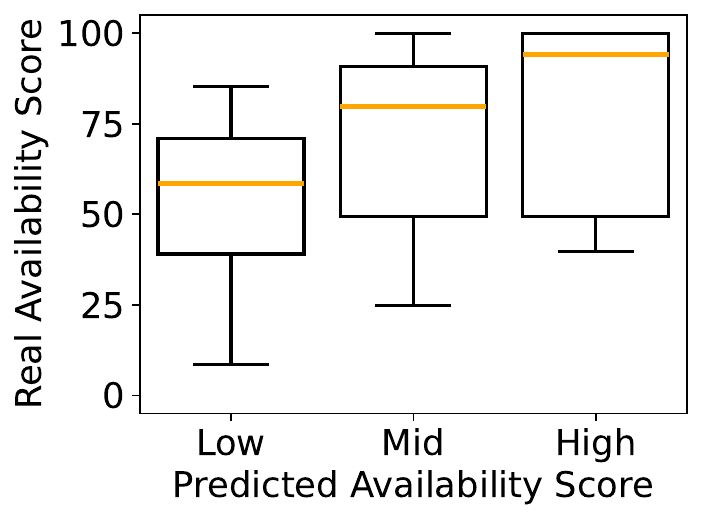}
        \label{fig:relationship-of-sps-score-and-real-score-fig}
    }\hfill
    \subfloat[Using vanilla SPS datasets ($T3$) without score calculation]{
        \includegraphics[width=0.47\columnwidth]{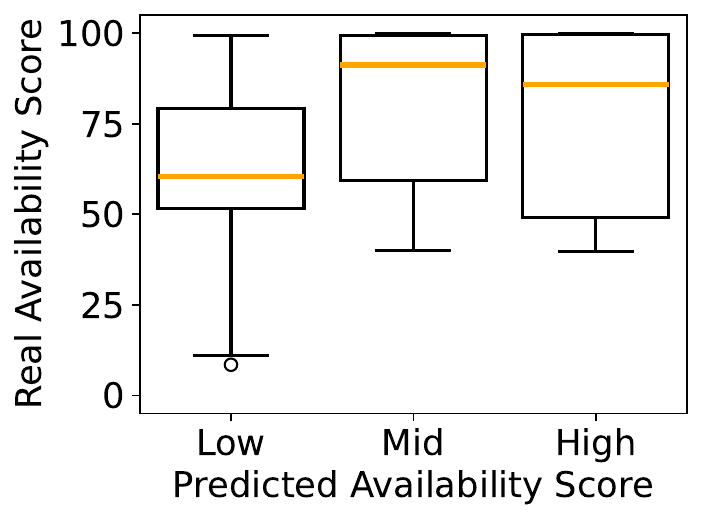}
        \label{fig:relationship-of-single-t3-and-real-score}
    }
    \caption{Superb real availability modeling capability of the proposed reliability scoring mechanism}
    \label{fig:real-predicted-availability-scores}
\end{figure}

\paragraph{Quantifying Effectiveness of the Availability Scoring Algorithm}
Figure~\ref{fig:real-predicted-availability-scores} compares the performance of these two predictors. The horizontal axis categorizes predicted availability score as Low ($<20$), Mid ($20–70$), and High ($>70$). The vertical axis shows the real availability score of instance types in each category based on the predicted scores. The proposed scoring heuristic (Figure~\ref{fig:relationship-of-sps-score-and-real-score-fig}) and the baseline (Figure~\ref{fig:relationship-of-single-t3-and-real-score}) show a positive correlation between their predicted scores and the real availability. However, a critical difference emerges in the \emph{Low} score category where the baseline predictor incorrectly assigns many stable instances low scores and thus exhibits poor recall in finding stable spot instances; 26.3\% of instances it categorized as \emph{Low} were, in fact, highly available. In contrast, the proposed heuristic demonstrates superior recall, with a misclassification error rate of 11.1\% in the same category.

This result confirms that the proposed availability scoring mechanism, which incorporates temporal characteristics like trend and volatility, provides a significantly more accurate and reliable model of real-world spot instance stability than a predictor based on a single, unprocessed metric.

After collecting the stability experiment data, we applied the Kaplan-Meier estimator (KME)~\cite{kaplan-meier} and the Cox proportional hazards model~\cite{cox-regression-model} to quantitatively evaluate spot instance survival times with the proposed availability score. These methods, commonly used to estimate survival rates or analyze variable impacts on survival~\cite{cox-example-breakaway, cox-example-cancer, cox-model-glaucoma-medications}, are widely applied in fields like medicine and business where the customer retention rate is important. Given their effectiveness in modeling lifetime data, they are well-suited for analyzing spot instance stability.

The hazard ratio of the Cox proportional model is calculated as follows.

\begin{equation}
h(t|x) = h_0(t) \exp((x - \bar{x})'\beta)
\end{equation}

$h(t|x)$ represents the hazard ratio at time $t$ given the availability score $x$, while $h_0(t)$ denotes the baseline hazard ratio without considering availability. $\beta$ is the regression coefficient, estimated using the log-likelihood function~\cite{log-likelihood-estimation}, to quantify the impact of availability on spot interruptions.

The results show a strong correlation between availability score and survival ratio ($P \leq 0.05$), with a hazard ratio of 0.9903 (95\% confidence interval: 0.9899–0.9907). Each 1-point increase in availability score reduces interruption risk by approximately 0.97\%, following $e^{-0.0097 \times \Delta x}$. At an availability score of 100, the risk decreases by about 62.1\% compared to a score of 0, confirming its effectiveness as a reliability indicator.

Next, to visualize the availability of multi-node spot instances based on the proposed availability score, we estimated their survival rate using the KME, calculated as follows:

\begin{equation}
\widehat S(t) = \prod\limits_{i:\ t_i\le t} \left(\frac{n_i - d_i}{n_i}\right)
\end{equation}

\begin{figure}[t]
    \centering
    \includegraphics[width=\columnwidth]{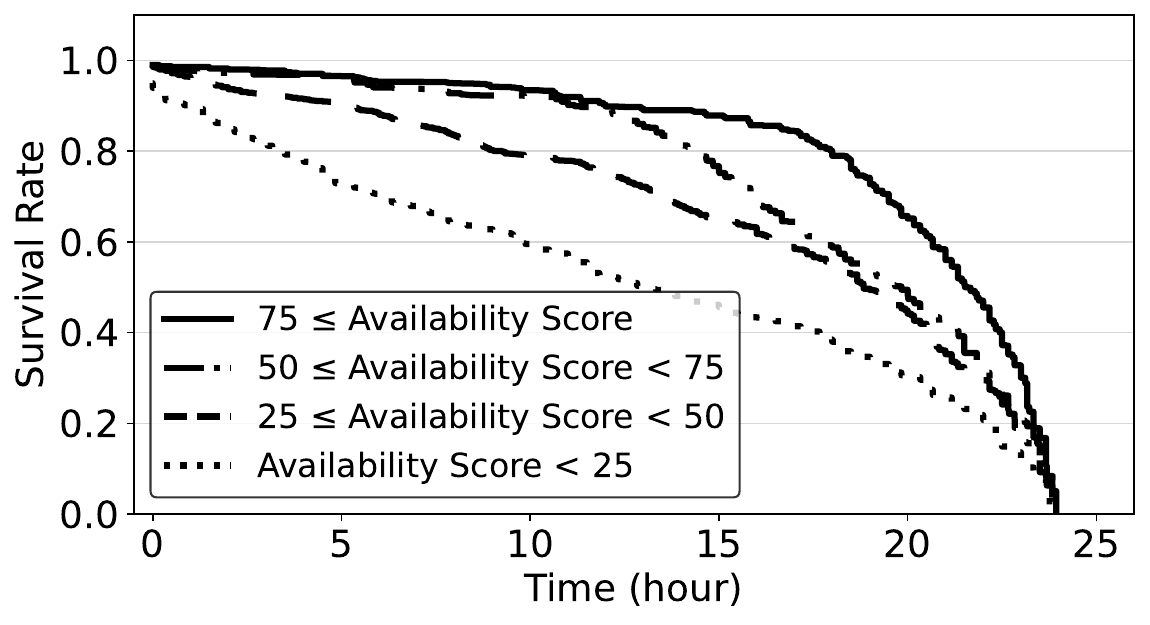}
    \caption{The survival rate of spot instances with a different maximum number of high availability nodes}
    \label{fig:multi-node-sps-survial-rate-kme}
\end{figure}

The survival function $\widehat S(t)$ represents the probability that a spot instance runs beyond time $t$. Here, $n_i$ is the number of running instances at time $t_i$, and $d_i$ is the number of interrupted instances. Figure~\ref{fig:multi-node-sps-survial-rate-kme} compares $\widehat S(t)$ across different availability score ranges. The x-axis represents runtime, while the y-axis shows the survival rate. Availability scores are grouped into four bins: solid lines indicate scores of 75 or higher, dashed-dotted lines represent scores between 50 and 75, and so on. Higher scores correspond to longer survival times—instances with scores below 25 have a median survival time of 13 hours, while those scoring 75+ last 21.6 hours. This confirms that the proposed availability score effectively enhances spot instance reliability.

\paragraph{Sensitivity of the Scaling Coefficient $\lambda$}
The availability score (Eq.~\ref{eq:sps-availability-score}) includes a scaling coefficient $\lambda$ that controls the magnitude of the trend and volatility adjustment. To validate the choice of $\lambda = 0.1$, a sensitivity analysis was conducted by sweeping $\lambda$ from $0.0$ to $1.0$ in increments of $0.1$. The accuracy was measured as the agreement between the predicted availability scores and the real availability scores derived from ground-truth interruption data, with improvements computed relative to the unadjusted baseline at $\lambda = 0.0$.

\begin{figure}[t]
    \centering
     \includegraphics[width=\columnwidth]{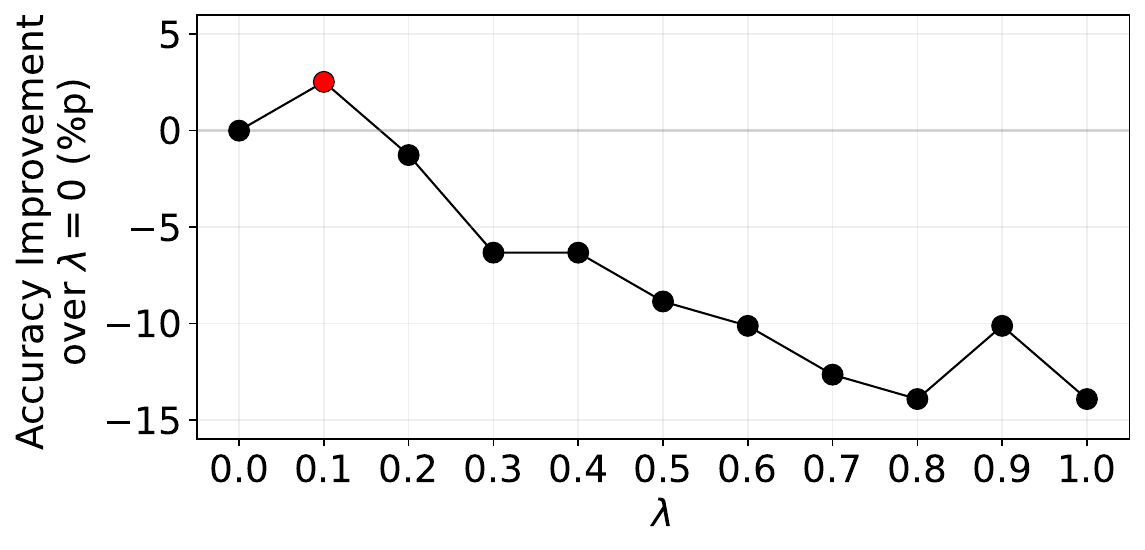}
    \caption{Sensitivity of the availability score to the scaling coefficient 
    $\lambda$.}
    \label{fig:lambda-sensitivity}
\end{figure}

Figure~\ref{fig:lambda-sensitivity} plots the accuracy improvement as a function of $\lambda$. The accuracy improvement reaches its peak at $\lambda = 0.1$ with an improvement of $+2.5$ percentage points over the baseline. For $\lambda \geq 0.2$, the adjustment over-amplifies estimation noise and the accuracy falls below the baseline across the remaining range. These results identify $\lambda = 0.1$ as the only operating point at which the trend and volatility adjustment improves the agreement between predicted and real availability scores, justifying its selection as the default coefficient in the proposed scoring model.

\paragraph{Impact of T3 Observation Period}
The length of the observation window used to compute the availability score from the $T3$ time-series directly affects score stability. To determine an appropriate window size, a sensitivity analysis was conducted by measuring $|\Delta AS|$, the absolute score difference between two consecutive window sizes when calculating the availability score, for the same base date.

\begin{figure}[t]
    \centering
    \includegraphics[width=\columnwidth]{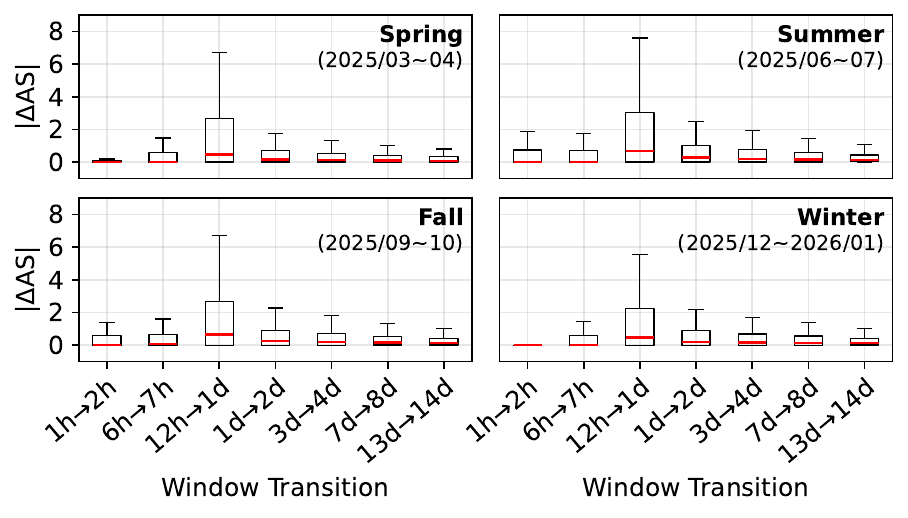}
    \caption{Distribution of $|\Delta AS|$ across window transitions for four seasons.}
    \label{fig:delta-as-window-transition}
\end{figure}

Figure~\ref{fig:delta-as-window-transition} presents the $|\Delta AS|$ distribution across seven window transitions over four seasonal periods from March 2025 to January 2026, each covering approximately 6,000 instance-transition observations. A consistent pattern emerges across all seasons: $|\Delta AS|$ peaks at the 12h$\rightarrow$1d transition, indicating that a full day of data is the critical threshold for capturing the daily cyclical patterns. Beyond one day, $|\Delta AS|$ drops sharply, and by the 7d$\rightarrow$8d transition, the median approaches near-zero. The consistency of this convergence behavior across all four seasons confirms that it is an inherent property of the $T3$ data rather than a seasonal artifact.

Based on this analysis, a seven-day observation window is adopted as the default for availability score computation. This duration captures weekly periodicity while remaining responsive to meaningful shifts in spot instance availability.

\paragraph{Validity of Using Only T3}
\begin{figure}[t]
  \centering
  \includegraphics[width=\columnwidth]{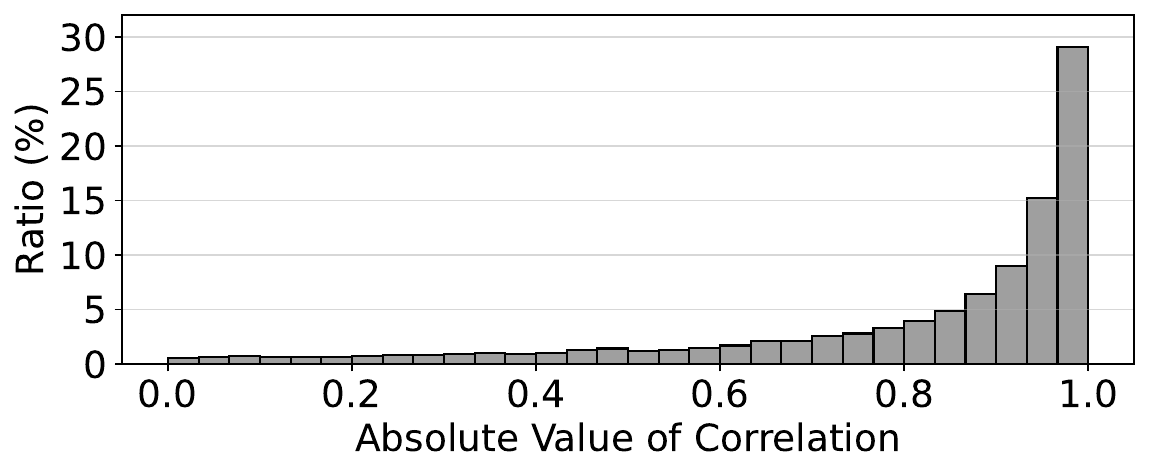}
  \caption{High correlations of $T3$ and $T2$}
  \label{fig:correlation-between-t3-and-t2-availability-score}
\end{figure}

To verify that incorporating $T2$ values when calculating the availability score is unnecessary, two availability scores were computed for each instance type over a one-week period (August 4--11, 2025): one from its $T3$ time-series and another from its $T2$ time-series. Figure~\ref{fig:correlation-between-t3-and-t2-availability-score} shows the distribution of Pearson correlation coefficients~\cite{pearson-correlation-coefficient} between the two scores. The distribution is heavily right-skewed, with approximately 25\% of instances exhibiting near-perfect correlation (coefficient close to 1.0), while instances with low correlation (below 0.6) are infrequent. This confirms that the values of $T3$ and $T2$ are highly synchronized, and an availability score derived solely from $T3$ is sufficient for assessing instance stability.

\begin{figure}[t]
    \centering
    \subfloat[Availability score]{
        \includegraphics[width=0.48\columnwidth]{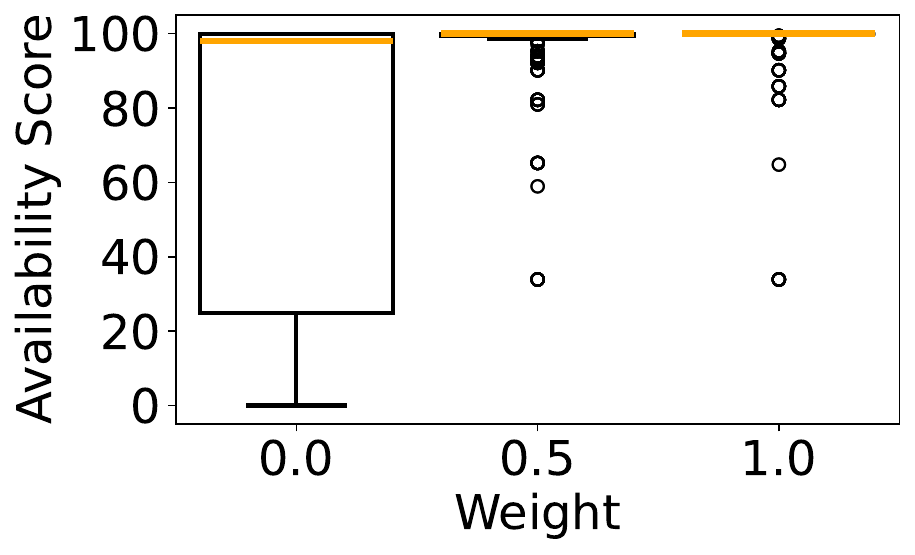}
        \label{fig:w-impact-to-availability-score}
    }
    \subfloat[Cost score]{
        \includegraphics[width=0.48\columnwidth]{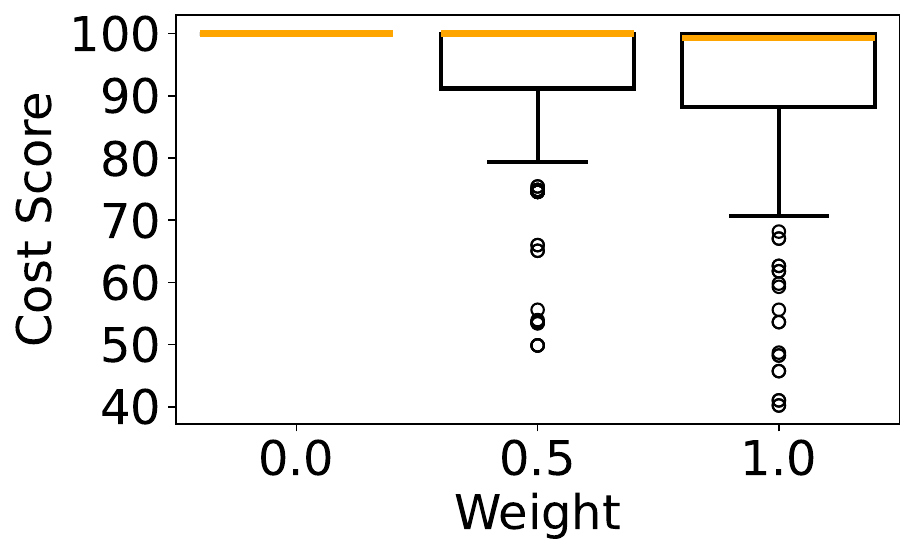}
        \label{fig:w-impact-to-cost-score}
    }
    \caption{Impact of weight ($W$) to score distribution}
    \label{fig:w-impact-to-score-distribution}
\end{figure}

\paragraph{Impact of Weight Parameter, $W$}
To analyze the effect of $W$ on ranking, 360 synthetic scenarios were constructed with varying vCPU (80--640) and memory (160--1280\,GB) requirements across different instance categories, families, and types. Availability and cost scores were computed using data from August 17--22, 2025, with $W$ set to 0.0, 0.5, and 1.0.

Figure~\ref{fig:w-impact-to-score-distribution} shows the score distributions for the top-ranked instances under each setting. At $W{=}0.0$ (cost-only), all selected instances achieve a perfect cost score of 100, but their availability scores are low and widely dispersed. At $W{=}1.0$ (availability-only), availability scores are near-perfect, but cost scores drop substantially. At $W{=}0.5$ (balanced), the engine achieves availability scores nearly identical to the $W{=}1.0$ case while maintaining high cost-efficiency, demonstrating that a balanced weight provides near-optimal availability with only a minor cost compromise. This justifies $W{=}0.5$ as the default setting.

\begin{table}[t]
\centering
\setlength{\tabcolsep}{3pt} 
\resizebox{0.48\textwidth}{!}{%
  \begin{tabular}{lcccccccc}
    \toprule
    \multirow{2}{*}{} & \multicolumn{4}{c}{vCPU} & \multicolumn{4}{c}{Memory} \\ 
    \cmidrule(lr){2-5} \cmidrule(lr){6-9}
            & 80   & 160  & 320  & 640  & 160  & 320  & 640  & 1280 \\ 
    \midrule
    Category & [2,4,5] & [2,3,5] & [3,4,8] & [4,4,7] & [2,4,5] & [2,5,8] & [3,3,4] & [4,4,7] \\
    Family   & [1,4,7] & [2,3,8] & [2,3,8] & [3,5,7] & [2,4,7] & [2,4,8] & [2,3,7] & [3,4,8] \\
    Types    & [3,3,3] & [3,4,4] & [4,4,4] & [4,4,4] & [3,3,3] & [3,4,4] & [3,4,4] & [4,4,4] \\
    \bottomrule
  \end{tabular}%
}
\caption{The number of minimum, median, and maximum types used in instance combinations for each scenario.}
\label{tab:number-of-instance-types-used-in-combinations}
\end{table}

\begin{figure}[t]
    \centering
    \subfloat[Instance Category]{
        \includegraphics[width=0.48\columnwidth]{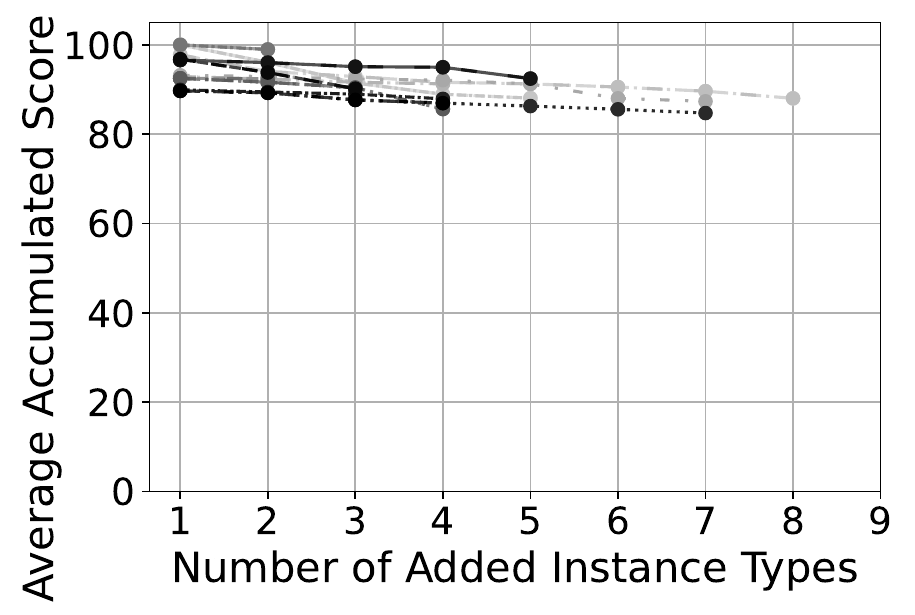}
        \label{fig:total-scores-added-node-instance-category}
    }
    \subfloat[Instance Family]{
        \includegraphics[width=0.48\columnwidth]{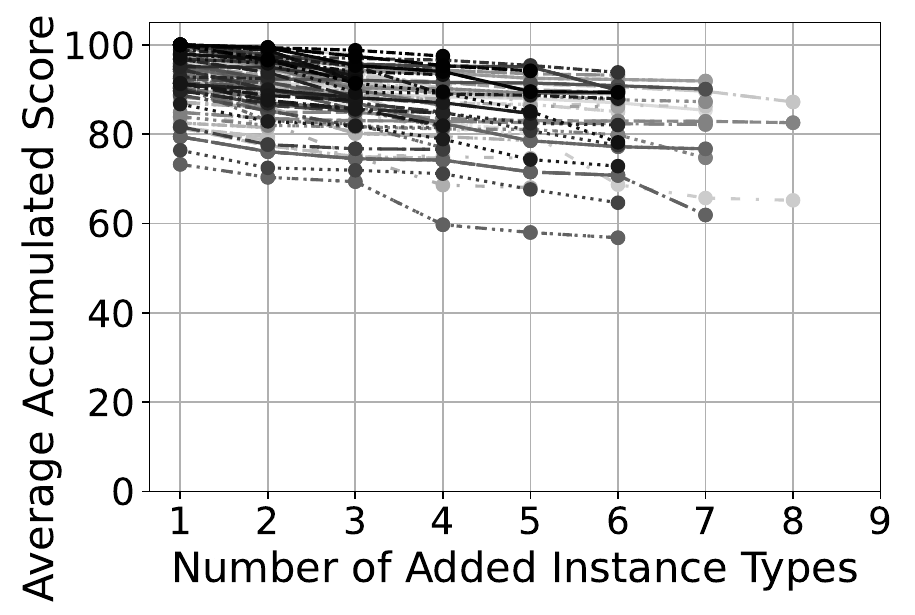}
        \label{fig:total-scores-added-node-instance-family}
    }
    \caption{Average total scores as we diversify a set of recommended instance types}
    \label{fig:total-scores-added-node}
\end{figure}

\paragraph{Effectiveness of Diversifying Instance Types} 
Relying on a single instance type for a large spot request increases interruption risk, motivating the proposed heuristic for constructing heterogeneous pools. Table~\ref{tab:number-of-instance-types-used-in-combinations} reports the [minimum, median, maximum] number of distinct instance types selected across scenarios with varying resource requirements (vCPU: 80--640; Memory: 160--1280\,GB) and three candidate pool scopes: broad (\textit{Category}), restricted (\textit{Family}), and narrow (\textit{Types}). The results show that the algorithm adaptively adjusts pool diversity based on both the scale of the request and the breadth of available candidates.

Figure~\ref{fig:total-scores-added-node} examines whether this diversification degrades recommendation quality. As instance types are progressively added to the pool, the average score declines only marginally in both broad (Figure~\ref{fig:total-scores-added-node-instance-category}) and restricted (Figure~\ref{fig:total-scores-added-node-instance-family}) candidate settings. This confirms that the heuristic's termination conditions stop diversification before lower-ranked additions substantially degrade pool quality, achieving increased resilience with minimal score compromise.

\subsubsection{Validation of Greedy-based Recommendation Approach}
\label{sec:greedy-vs-ilp-evaluation}
To validate the effectiveness of the proposed greedy heuristic for recommendation, we compare it against an ILP formulation that jointly maximizes the total score and instance-type diversity.

\paragraph{ILP Formulation}
To construct a reasonable baseline for comparison, we formulate an ILP that encodes both score maximization and type diversification as a single objective. While this formulation is not intended as a general-purpose optimal solver, it provides a useful reference point for assessing the quality gap of the greedy heuristic.

For each candidate instance $i$, an integer variable $x_i$ denotes the number of allocated nodes, and a binary variable $z_i$ indicates whether instance type $i$ is selected. The objective maximizes $\sum_i S_i \cdot \text{CPU}_i \cdot x_i + \gamma \sum_i z_i$, where the first term measures the vCPU-weighted pool quality and the second term rewards type diversity with coefficient $\gamma$. Linking constraints enforce $z_i = 1$ whenever $x_i > 0$, and a resource constraint ensures $R_{\text{req}} \le \sum_i \text{CPU}_i \cdot x_i \le R_{\text{req}} + 1$ to minimize over-provisioning. The binary variables expand the search space exponentially with the number of candidates. PuLP~\cite{pulp} with the CBC solver is used without a time limit.

\paragraph{Setup and Results}
Six days of SPS data (August 17--22, 2025) are used with a fixed requirement of 160 vCPU and $W{=}0.5$. The candidate set is scaled by progressively adding AWS regions from 1 to 17, growing the candidate instance types from 808 to 33,279. Table~\ref{tab:greedy-vs-ilp-runtime} reports execution time and total pool score at four representative scales. The user requirement is fixed at 160 vCPU, and the availability-cost weight $W$ is set to 0.5. PuLP with the CBC solver is used for ILP execution.

The greedy heuristic runs in approximately 2--3\,ms regardless of scale, owing to the early termination after exploring only the top-ranked candidates. The ILP solver exhibits rapidly accelerating execution time, which is from 154ms at 808 candidates to 24.7 seconds at 33,279 candidates, a 160$\times$ increase in runtime for a 41$\times$ increase in candidate count. At full scale, the ILP is over 8,000$\times$ slower than the greedy heuristic.

In terms of solution quality, the score gap is at most 0.3\% at full scale. At intermediate scales (4 and 10 regions), the greedy heuristic occasionally matches or slightly exceeds the ILP score; this occurs because the two methods optimize diversity differently, where the greedy heuristic uses score-proportional allocation while the ILP uses a linear diversity bonus, so neither strictly dominates the other in all cases. Overall, the greedy heuristic achieves comparable solution quality at a fraction of the computational cost, confirming its suitability for real-time recommendation over large candidate spaces.

\begin{table}[t]
\caption{Execution time and score comparison between the greedy heuristic and ILP across candidate space scales}
\label{tab:greedy-vs-ilp-runtime}
\centering
\footnotesize
\begin{tabular}{rrrrrr}
\toprule
& & \multicolumn{2}{c}{\textbf{Time (ms)}} & \multicolumn{2}{c}{\textbf{Sum of Score}} \\
\cmidrule(lr){3-4} \cmidrule(lr){5-6}
\textbf{Regions} & \textbf{Candidates} & \textbf{Greedy} & \textbf{ILP} & \textbf{Greedy} & \textbf{ILP} \\
\midrule
1  & 808    & 2.3 & 154    & 7,415 & 8,059 \\
4  & 4,584  & 2.3 & 607    & 8,072 & 8,061 \\
10 & 14,298 & 2.4 & 2,539  & 8,001 & 8,046 \\
17 & 33,279 & 3.0 & 24,725 & 8,000 & 8,024 \\
\bottomrule
\end{tabular}
\end{table}

\begin{figure*}[t]
    \centering
    \includegraphics[width=\columnwidth]{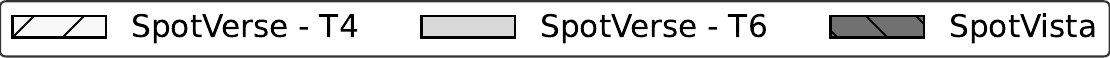}\\
    \subfloat[Incurred cost per hour]{
        \includegraphics[width=\columnwidth]{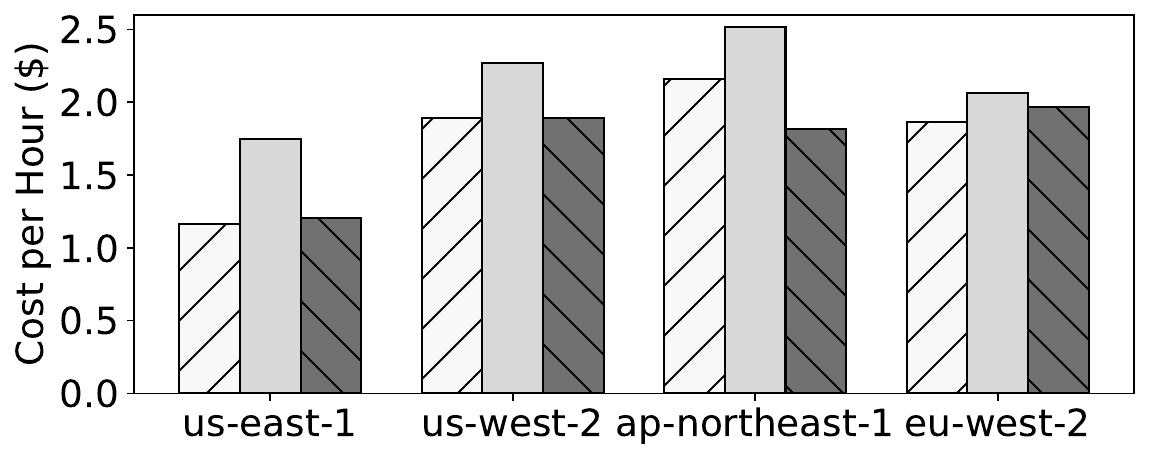}
        \label{fig:SpotVista-spotverse-cost}
    }
    \subfloat[Spot instance available time percentage]{
        \includegraphics[width=\columnwidth]{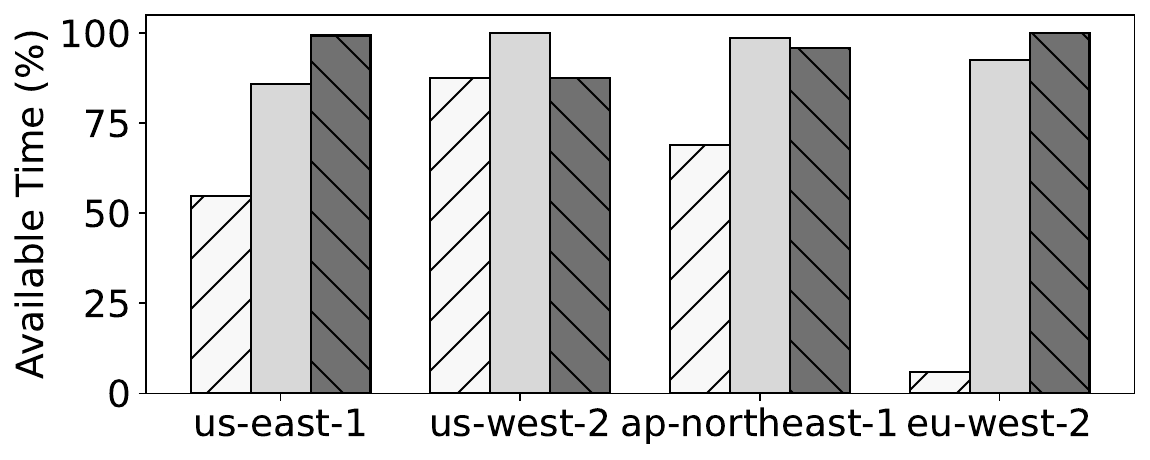}
        \label{fig:SpotVista-spotverse-availability}
    }
    \caption{Comparing the spot instance recommendation quality of SpotVista with SpotVerse}
    \label{fig:SpotVista-spotverse-compare}
\end{figure*}

\begin{figure*}[t]
    \centering
    \subfloat[Cost savings ratio]{
        \includegraphics[width=\columnwidth]{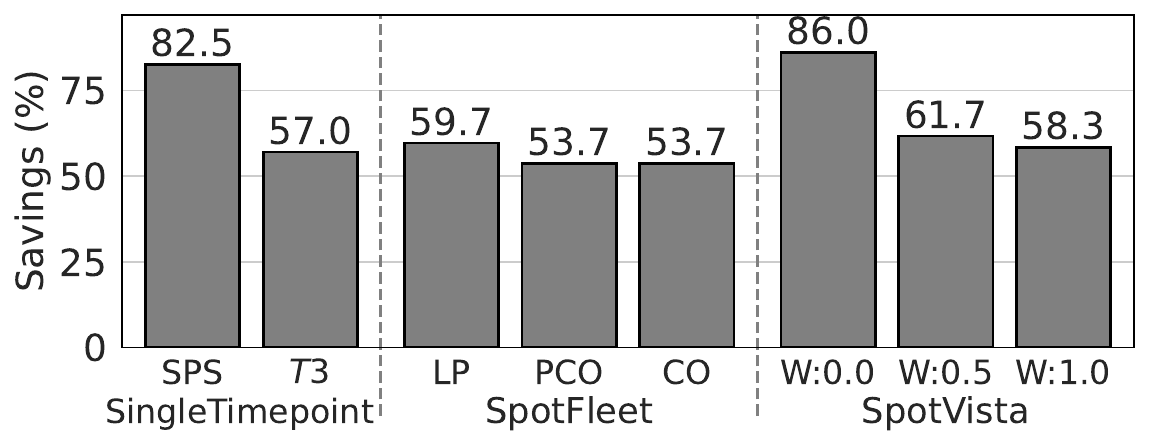}
        \label{fig:compare-cost}
    }
    \subfloat[Availability]{
        \includegraphics[width=\columnwidth]{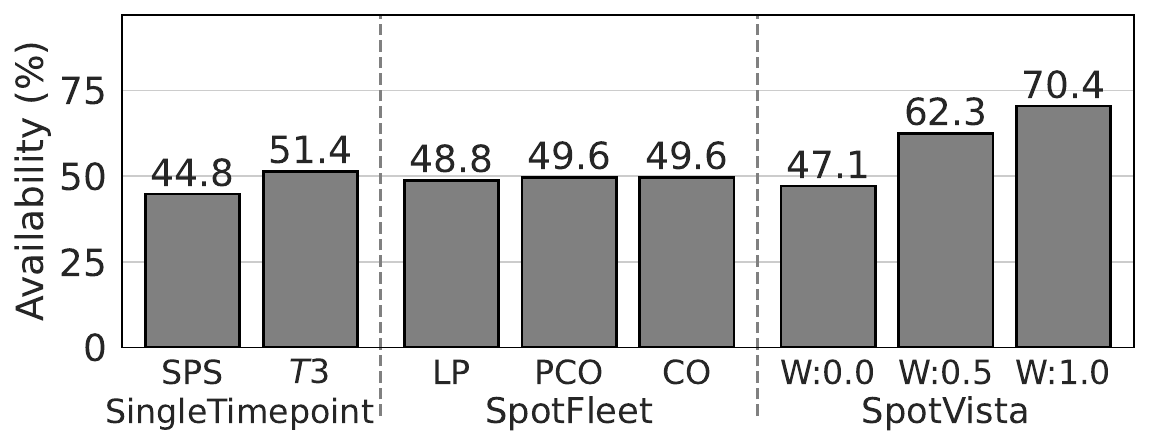}
        \label{fig:compare-availability}
    }
    \caption{Comparing SpotVista with AWS SpotFleet and simple approaches without using historical dataset}
    \label{fig:compare-availability-cost}
\end{figure*}

\subsection{Effectiveness of Instance Recommendation}
To address \textbf{RQ-4}, the effectiveness of SpotVista is evaluated against the state-of-the-art multi-region spot instance recommendation system, SpotVerse~\cite{spotverse}, and a public commercial service, AWS SpotFleet~\cite{spotfleet} in terms of cost and reliability.

\paragraph{Experimental Setup}
SpotVerse recommends stable, cost-efficient spot instances in a multi-region setup, using SPS and Interruption-Free (IF) scores. The IF score represents the ratio of interruptions over the past 30 days, ranging from 1 to 3. SpotVerse sums SPS and IF scores, filtering instances with a total score above a threshold (default $T=4$). It then selects the cheapest instance among candidates. To observe cases where availability is prioritized over cost, we also tested with $T=6$. For a fair comparison, we conducted experiments in four major AWS regions (us-west-2, us-east-1, eu-west-2, ap-northeast-1) following SpotVerse’s setup; requesting the amount of compute and memory resources of 40 $\times$ \emph{m5.xlarge} instances per region with thresholds $T=4$ and $T=6$. The interruption modeling experiment~\cite{cant-be-late} ran for 24 hours from Feb. 3 to Feb. 5, 2025. While SpotVista normally recommends a mix of instance types to enhance stability, we constrained it to a single instance type per experiment to align with SpotVerse’s methodology that does not support diversifying instance types.

\paragraph{Comparing with SpotVerse} Figure~\ref{fig:SpotVista-spotverse-compare} presents the comparative results for incurred hourly cost and instance availability. The analysis reveals that SpotVista consistently outperforms both SpotVerse configurations.

In terms of cost (Figure~\ref{fig:SpotVista-spotverse-cost}), SpotVista is the most efficient approach, reducing costs by 2.84\% compared to SpotVerse-T4 and 20.02\% compared to the more expensive SpotVerse-T6. This is because SpotVista's balanced scoring model can identify cost-efficient options that SpotVerse's availability-first filtering mechanism overlooks.

Figure~\ref{fig:SpotVista-spotverse-availability} compares the percentage of time during which spot instances remained stable. SpotVerse-T6 places greater emphasis on spot instance reliability, resulting in significantly higher stability compared to SpotVerse-T4. The availability performance of SpotVista closely aligns with that of SpotVerse-T6. Overall, SpotVista achieves 1.53\% higher availability than SpotVerse-T6 and 81.28\% higher than SpotVerse-T4.

In summary, SpotVista simultaneously achieves superior cost-efficiency and state-of-the-art availability, outperforming the existing state-of-the-art in both dimensions. Comparisons with older spot price prediction strategies~\cite{spot-price-prediction, spot-price-rf, spot-bidding-infocom, spot-price-prediction-dnn} are excluded, as recent studies have shown their reduced effectiveness due to changes in cloud vendor pricing policies~\cite{spot-price-change-2017, spot-price-policy-change-2017-irwin}.

\paragraph{Comparing with AWS SpotFleet} Next, we compare SpotVista with AWS SpotFleet~\cite{spotfleet}, which enables users to configure a pool of spot instances using a launch template and adjust allocation strategies for cost and stability. SpotFleet supports three strategies: \emph{Lowest Price} (LP), \emph{Capacity Optimized} (CO), and \emph{Price-Capacity Optimized} (PCO). SpotVista can emulate these strategies by adjusting $W$: $W=0.0$ for LP, $W=1.0$ for CO, and $W=0.5$ for PCO. We evaluate all configurations.

Due to SpotFleet’s regional constraints, experiments were conducted in \emph{us-east-1}, one of the largest AWS regions. Additionally, we compared performance against a naive approach that selects multi-node spot instances based solely on single-node SPS and $T3$ values at the request time, ignoring temporal effects. When multiple instances had the same highest values, we selected the lowest-priced one. For evaluation, we requested 50 requests every 10 minutes over a 24-hour period and recorded the success rate. 

Figure~\ref{fig:compare-availability-cost} presents a bar plot comparing availability and cost across different instance selection strategies. The horizontal axis represents the strategies, while the vertical axis shows cost savings (Figure~\ref{fig:compare-cost}) and availability (Figure~\ref{fig:compare-availability}). In Figure~\ref{fig:compare-cost}, cost savings increase as SpotVista’s weight ($W$) decreases. SpotFleet's lowest price (LP) strategy achieves slightly higher savings than its other strategies but is 30.6\% lower than SpotVista at $W=0$. The single time-point SPS approach also achieves high cost savings by selecting the cheapest instances with an SPS of three but still falls short of SpotVista’s cost efficiency.

Figure~\ref{fig:compare-availability} shows that SpotVista improves availability as $W$ increases. SpotFleet's CO and PCO strategies provide slightly better availability (0.8\%) than LP but are 29.5\% lower than SpotVista at $W=1$. CO and PCO perform identically because SpotFleet makes the same recommendations, and their availability remains similar to LP, failing to meet expectations. All single time-point strategies show lower availability than SpotVista at $W=0.5$ and $W=1.0$.

Overall, compared to AWS SpotFleet, SpotVista improves availability by over 20\% while maintaining similar cost savings. Additionally, when availability is comparable, SpotVista achieves over 25\% more cost savings, demonstrating its effectiveness and practicality.

\section{Related Work}
\textbf{Modeling and Utilization of Spot Instance Datasets}: Since the introduction of spot instances, many attempts have been made to correlate spot instance price datasets with interruption risks to minimize the likelihood of instance termination~\cite{draft-spot-instance-guarantee-from-spot-price, alibaba-spot-instance, stat-analysis-spot-price, spot-analysis-javadi, spot-instance-analysis, spot-price-by-location, spot-instance-for-hpc}. For example, Ali-Eldin et al. proposed a heuristic of the deployment of web servers based on spot prices and analysis~\cite{spotweb}, while Lee et al. introduced DeepSpotCloud~\cite{deepspotcloud}, which uses GPU spot instances across global regions for DNN training tasks. In big data processing, SeeSpotRun~\cite{see-spot-run} suggested methods for efficiently using spot instances in a Hadoop~\cite{hadoop} cluster, while Flint~\cite{flint} and Tr-Spark~\cite{tr-spark} were proposed for Apache Spark~\cite{spark}. Fabra et al.~\cite{spot-price-prediction-dnn} proposed a DNN model to predict spot instance prices to enhance its reliability. However, these previous works relied on historical spot price datasets, which became obsolete after the operational changes~\cite{spot-price-change-2017, spot-price-policy-change-2017-irwin}. This paper presents a new approach that does not depend on spot price datasets, enabling the selection of cost-efficient and stable multiple spot instances in a different way.

\textbf{Spot Instance Interruption Analysis}: Pham et al.\cite{spot-instance-interrupt-check-cloud-2018}, Lee et al.\cite{spotlake-iiswc}, and Kim et al.\cite{interrupt-visible-www} conducted experiments on AWS spot instances, analyzing interruption patterns. In the case of Azure, Yang et al.\cite{azure-spot-prediction} proposed an interruption prediction model based on the Transformer architecture, using Azure's internal interruption tracking data. For GCP, Haugerud et al.\cite{gcp-spot-vm-batch-cost-reduction} and Kadupitiya et al.\cite{scispot} modeled spot instance interruptions. Additionally, Yang et al.~\cite{skypilot} performed interruption experiments in multi-cloud environments. Our proposed approach is complementary to previous research, as it leverages a new type of spot instance dataset to improve reliability.

\textbf{Utilization of Multi-Node Spot Instances}: As the scale for computing resources increases for various applications, such as deep learning and big data processing, several approaches have been proposed to reliably utilize multiple spot instance nodes, focusing on reducing training costs while handling interruptions and improving training throughput~\cite{bamboo-nsdi, parcae-nsdi, oobleck-sosp, varuna-eurosys, deepvm-ccgrid}. We expect that the algorithm proposed in this work will increase the utility of these previous research outcomes. Yang et al.\cite{snape-azure-spot-mixture} and Xu et al.\cite{boss-inmemory-infocom} proposed methods to build cost-efficient and reliable clusters by mixing on-demand and spot instances. Sharma et al.\cite{ExoSphere} and Harlap et al.\cite{tributary-atc18} introduced frameworks that balance cost and reliability for spot instance clusters. While existing work relies on internal datasets or spot price datasets, to the best of our knowledge, this paper is the first to quantitatively model the availability of large-scale spot instances using public datasets other than price.

\section{Discussion}
\label{sec:discussion}
\paragraph{Generalization to Other Cloud Vendors}
The methodology of SpotVista is vendor-agnostic, but its applicability requires two prerequisites. A queryable availability indicator analogous to AWS SPS must be exposed, and the collected data must support reliable historical pattern-based scoring. Azure has recently begun offering an SPS API~\cite{azure-spot-placement-score}, and the MSTL analysis in Section~\ref{sec:multi-node-dataset-analysis} shows that Azure $T3$ exhibits substantially weaker seasonality and higher amplitude variability than AWS. In addition, the data collected from the Azure API contains missing and inconsistent responses, which hinders continuous time-series collection. A complete extension would require additional preprocessing such as missing-value imputation, together with an adapted scoring model that accounts for weaker periodic structure. GCP and Alibaba do not currently expose a public availability score API, but the proposed methodology can be extended to these platforms once comparable interfaces become available.

\paragraph{Reactive Adjustment after Deployment}
SpotVista currently operates as a one-shot recommendation engine and does not monitor spot instance availability or cost score changes after deployment. A natural extension is to integrate SpotVista as an availability signal provider for workload-agnostic cluster managers such as SkyPilot~\cite{skypilot}, enabling continuous rebalancing throughout the workload lifetime. Such an extension would require a reactive decision loop that periodically reevaluates the active pool against updated availability signals, which we consider a promising direction for future research.

\section{Conclusion}
\label{sec:conclusion}
This paper addressed the critical challenge of provisioning reliable multi-node workloads on volatile cloud spot instances, a problem for which single-node availability metrics are inadequate. To solve this challenge, we proposed SpotVista whose key contributions include an efficient, sampling-based heuristic for collecting large-scale multi-node availability data, a new scoring model that quantifies instance stability by analyzing temporal patterns, and a recommendation engine that constructs diverse, cost-efficient, and reliable instance pools.

Through extensive real-world experiments, SpotVista demonstrated significant performance improvement over existing solutions. Compared to the state-of-the-art research system, SpotVerse~\cite{spotverse}, it achieved 81.28\% greater reliability and 2.84\% higher cost-efficiency. Furthermore, it provided 21.6\% higher stability than the commercially available AWS SpotFleet service, underscoring its practical effectiveness.

To benefit the research community and cloud practitioners, the collected multi-node availability dataset and the recommendation engine have been made publicly available through a web service. Future work will focus on extending the system to support dynamic, post-recommendation adjustments and expanding compatibility to other major cloud vendors.

\section*{Acknowledgements}
This work was supported by Institute of Information \& communications Technology Planning \& Evaluation(IITP) grant funded by the Korea government(MSIT) (RS-2022-00144309 \& RS-2025-25441560 \& RS-2026-25492200)


\bibliographystyle{elsarticle-num}
\bibliography{revision_spotvista_manuscript} 

\end{document}